\documentclass[article,5p,twocolumn]{elsarticle}
\usepackage{graphicx,latexsym}
\usepackage{amssymb,amsmath,bm}
\usepackage{subfigure}
\usepackage{braket}

\newcommand{\brames}[1]{( #1|}
\newcommand{\ketmes}[1]{|#1)}

\usepackage{hyperref}
\hypersetup{
    pdfnewwindow=true,       % links in new window
    colorlinks=true,         % false: boxed links; true: colored links
    linkcolor=blue,          % color of internal links
    citecolor=blue,          % color of links to bibliography
    filecolor=magenta,       % color of file links
    urlcolor=black           % color of external links
}

\def\eq#1{Eq.\ (\ref{#1})}
\def\mb#1{\mbox{\boldmath$#1$}}
\def\fig#1{Fig.\ \ref{#1}}

\begin{document}

\begin{frontmatter}

%-----------------------------------------------------------------
\title{Impact of a circularly polarized cavity photon field\\
on the charge and spin flow through an Aharonov-Casher ring}

\author[a1]{Thorsten Arnold\corref{cor1}}
\ead{tla1@hi.is}
\address[a1]{Science Institute, University of Iceland, Dunhaga 3,
        IS-107 Reykjavik, Iceland}
\author[a2]{Chi-Shung Tang}
\ead{cstang@nuu.edu.tw}
\address[a2]{Department of Mechanical Engineering,
        National United University,
        1, Lienda, Miaoli 36003, Taiwan}
\author[a3]{Andrei Manolescu}
\ead{manoles@hr.is}
\address[a3]{School of Science and Engineering, Reykjavik University,
        Menntavegur 1, IS-101 Reykjavik, Iceland}
\author[a1]{Vidar Gudmundsson}
\ead{vidar@hi.is}
\cortext[cor1]{Corresponding author (phone: +354 5254181)}
%
%----------------------------------------------------------------

\begin{abstract}
 \ We explore the influence of a circularly polarized cavity photon field on the transport properties
 of a finite-width ring, in which the electrons are subject to spin-orbit and Coulomb interaction.
 The quantum ring is embedded in an electromagnetic cavity 
 and described by ``exact'' numerical diagonalization.
 We study the case that the cavity photon field is circularly polarized
 and compare it to the linearly polarized case.
 The quantum device is moreover coupled to external, electrically biased leads. The
 time propagation in the transient regime is described
 by a non-Markovian generalized master equation.
 We find that the spin polarization and spin photocurrents of the quantum ring
 are largest for circularly
 polarized photon field and destructive Aharonov-Casher (AC) phase interference.
 The charge current suppression dip due to the destructive AC phase becomes threefold under
 the circularly polarized photon field as the interaction of the electrons' angular momentum and
 spin angular momentum of light causes many-body level splitting
 leading to three many-body level crossing locations instead of one.
 The circular charge current inside the ring, which is 
 induced by the circularly polarized photon field, is found to be suppressed in
 a much wider range around the destructive AC phase
 than the lead-device-lead charge current.
 The charge current can be directed through one of the two ring arms
 with the help of the circularly polarized photon field, but is superimposed
 by vortices of smaller scale.
 Unlike the charge photocurrent, the flow direction of the spin photocurrent
 is found to be independent of the
 handedness of the circularly polarized photon field.
\end{abstract}

\begin{keyword}
Cavity quantum electrodynamics \sep Electronic transport \sep Aharonov-Casher effect \sep Quantum ring 
\sep Spin-orbit coupling \sep Circularly polarized photon field
\PACS 78.67.-n \sep 73.23.-b \sep 85.35.Ds \sep 71.70.Ej
\end{keyword}

\end{frontmatter}

\section{Introduction}

Quantum rings are interferometers with unique properties
owing to their rotational symmetric geometry.
Because of their non-trivially connected topology, a variety of geometrical phases
can be observed~\cite{PhysRev.115.485, PhysRevLett.53.319, PhysRevLett.58.1593, Berry_phase},
which can be tuned via the magnetic flux through the ring
in case of the Aharonov-Bohm (AB) phase,
or the strength of the spin-orbit interaction (SOI)
in case of the Aharonov-Casher (AC) phase.
Furthermore, the rotational symmetry of the ring resembles the characteristics
of a circularly polarized photon field suggesting a strong light-matter interaction
between single photons and the ring electrons.
Circularly polarized light emission~\cite{PhysRevA.78.051804} and absorption~\cite{3336122820080714}
has been studied for quantum rings. Moreover, circularly polarized light has been used to generate 
persistent charge currents in quantum wells~\cite{PhysRevLett.86.4358} and quantum rings~\cite{Quinteiro:09, PhysRevB.72.245331, PhysRevLett.107.106802, PhysRevB.87.245437}.
The basic principle behind this is a change of the orbital angular momentum of the electrons
in the quantum ring by the absorption or emission of a photon leading to the circular charge transport.
We would like to mention that one can improve over circularly polarized light to 
achieve optimal optical control for a finite-width quantum ring~\cite{Rasanen07:157404}.
Rather than trying to optimize quantum transitions, we are focusing here 
on various interesting effects that a circularly polarized photon field has on quantum rings
of spin-orbit and Coulomb interacting electrons.

The transport properties of magnetic-flux threaded rings,
which are connected to two leads
have been investigated in detail~\cite{Szafran05:165301,Pichugin56:9662}.
The conductance of the ring shows characteristic oscillations with period $\Phi_0=hc/e$, called
AB oscillations,
which were measured for the first time in 1985~\cite{Webb.54.2696}.
The electrons' spin is not only interacting with a magnetic field via the Zeeman interaction,
but also with an electric field via a so-called effective magnetic field stemming from special relativity~\cite{1367-2630-9-9-341}.
The interaction between the spin and the electronic motion in, 
for example, the electric field, is called the Rashba SOI~\cite{0022-3719-17-33-015}, which leads
to the AC effect.
Experimentally, the strength of the Rashba interaction
can be varied by changing the magnitude of the electric field 
when it is oriented parallel to the central axis of the quantum ring.
Another type of SOI is the Dresselhaus interaction~\cite{PhysRev.100.580},
leading to the AC effect as well.
The combined effects of SOI
and an applied magnetic field on
the electronic transport through quantum rings connected to leads
have been addressed in several studies~\cite{PhysRevB.69.235310,0256-307X-21-11-006,PhysRevLett.70.343,PhysRevB.55.10631,PhysRevB.72.165336}.
In this work, we use a small magnetic field outside the AB regime
and a tunable Rashba
or Dresselhaus SOI up to a strength corresponding to
an AC phase difference $\Delta\Phi \approx 3\pi$ and use cavity quantum electrodynamics
to describe the interaction of the electronic system with a
circularly polarized photon field in a cavity.

While the magnetic flux through the ring causes only equilibrium persistent \textit{charge} currents~\cite{PhysRevB.37.6050,Tan99:5626}, SOI can also induce equilibrium persistent
\textit{spin} currents~\cite{PhysRevLett.70.1678,PhysRevB.51.13441}.
Dynamical spin currents can be obtained by two asymmetric electromagnetic pulses~\cite{0953-8984-21-14-145801}.
Optical control of the spin current 
can be achieved by a non-adiabatic, two-component laser pulse~\cite{PhysRevB.83.155427}.
The persistent spin current is in general not conserved~\cite{PhysRevB.77.035327}.
Proposals to measure persistent spin currents by the induced mechanical torque~\cite{PhysRevLett.99.266602} 
or the induced electric field~\cite{PhysRevB.77.035327} exist.

Quantum systems embedded in an electromagnetic cavity have become one of the most promising
devices for quantum information processing applications~\cite{0034-4885-69-5-R02, 0953-4075-38-9-007, Physica_E_47_17}. 
We are considering here the influence
of the cavity photons on the spin polarization of the quantum ring
and on the transient charge and spin transport inside and into and out of the ring.
We treat the electron-photon interaction by using exact numerical diagonalization
including many levels~\cite{1367-2630-14-1-013036}, 
i.e. beyond a two-level Jaynes-Cummings model
or the rotating wave approximation and higher order corrections of it~\cite{1443594,PhysRevLett.98.013601,Sornborger04:052315}.
The electronic transport through a quantum system 
that is strongly coupled to leads
has been investigated for linearly polarized electromagnetic fields~\cite{Tang99:1830,Zhou17:6663,Jung85:023420,Tang00:127,Zhou68:155309}.
For a weak coupling between the system and the leads,
the Markovian approximation,
which neglects memory effects in the system, can be used~\cite{Spohn53:569,Gurvitz96:15932,Kampen01:00,Harbola06:235309}.
To describe a stronger transient system-lead coupling,
we use a non-Markovian generalized master equation~\cite{PhysRevLett.72.1076,Braggio06:026805,Moldoveanu09:073019}
in a time-convolutionless form~\cite{PhysRevA.59.1633,PhysRevB.87.035314},
which involves energy-dependent coupling elements.
We used this type of master equations earlier to explore the
interplay of linearly polarized cavity photons and topological phases of quantum rings
for the AB~\cite{PhysRevB.87.035314} and AC~\cite{2013arXiv1310.5870A} phase.
The influence of a quantized cavity photon mode of \textit{circularly} polarized light
on the time-dependent transport of spin-orbit and Coulomb interacting
electrons under non-equilibrium conditions 
through a topologically nontrivial broad ring geometry, which is connected to leads
has not yet been explored beyond the Markovian approximation. 
We note that we can compare our results to the analytic results
for a one-dimensional quantum ring with
Rashba or Dresselhaus SOI~\cite{2013arXiv1310.5870A}.

The paper is organized as follows. In Sec.\ \ref{sec2}, we describe the Hamiltonian 
for the central quantum ring system including SOI,
which is embedded in a photon cavity and our time-dependent generalized master equation formalism
for the transient coupling to semi-infinite leads. 
In addition, various transport quantities that are shown
in Secs.\ \ref{sec3} and \ref{sec4}, are here defined.
Sec.\ \ref{sec3} shows the numerical results concerning the spin polarization of the quantum ring,
both for linear and for circular polarization.
Sec.\ \ref{sec4} is devoted to the transport of charge and spin. First, 
the non-local currents (lead-ring currents) between the leads and the system are discussed.
Second, the change of the local currents inside the quantum ring due to the photon cavity 
(photocurrent) is considered. 
Finally, conclusions will be drawn in Sec.\ \ref{sec5}.

\section{Theory, model and definitions} \label{sec2}

Here, we describe the Hamiltonian of the central system including the potential used to model 
the quantum ring, the Hamiltonian for the leads and the time evolution of the whole system
described by a non-Markovian master equation. Furthermore, we define various quantities
describing the transient charge and spin transport and their accumulation
in the quantum ring. and the parameters used for the numerical results.

\subsection{Central system Hamiltonian}

The time-evolution of the closed many-body (MB) system 
composed of the interacting electrons and photons 
relative to the initial time $t=0$,
\begin{equation}
 \hat{U}_{S}(t)=\exp\left(-\frac{i}{\hbar}\hat{H}_{S}t\right),
\end{equation}
is governed by the MB system Hamiltonian
\begin{eqnarray}
\hat{H}_{S}&=&\int d^2 r\; \hat{\mathbf{\Psi}}^{\dagger}(\mathbf{r})\left[\left(\frac{\hat{\mathbf{p}}^2}{2m^{*}} +V_S(\mathbf{r})\right) + H_{Z}\nonumber \right. \\
 &&+\left. \hat{H}_{R}(\mathbf{r})+\hat{H}_{D}(\mathbf{r})\right]\hat{\mathbf{\Psi}}(\mathbf{r})+\hat{H}_{ee}+\hbar\omega \hat{a}^{\dagger}\hat{a}, \label{H^S}
\end{eqnarray}
with the spinor
\begin{equation}
 \hat{\mathbf{\Psi}}(\mathbf{r})=
 \left( \begin{array}{c} \hat{\Psi}(\uparrow,\mathbf{r}) \\ \hat{\Psi}(\downarrow,\mathbf{r}) \end{array} \right)
\end{equation}
and
\begin{equation}
 \hat{\mathbf{\Psi}}^{\dagger}(\mathbf{r})=
 \left(\begin{array}{cc} \hat{\Psi}^{\dagger}(\uparrow,\mathbf{r}), & \hat{\Psi}^{\dagger}(\downarrow,\mathbf{r}) \end{array}\right), \label{conj_FOS}
\end{equation}
where
\begin{equation}
 \hat{\Psi}(x)=\sum_{a}\psi_{a}^{S}(x)\hat{C}_{a}
\end{equation} 
is the field operator with $x\equiv (\mathbf{r},\sigma)$, $\sigma \in \{ \uparrow,\downarrow \}$ and the annihilation operator, $\hat{C}_{a}$,
for the single-electron state (SES) $\psi_a^S(x)$ in the central system.
The SES $\psi_a^S(x)$ is the eigenstate labeled by $a$
of the Hamiltonian $\hat{H}_{S}-\hat{H}_{ee}-\hbar\omega \hat{a}^{\dagger}\hat{a}$
when we set the photonic part of the vector potential $\hat{\mathbf{A}}^{\mathrm{ph}}(\mathbf{r})$
in the momentum operator,
\begin{equation}
 \hat{\mathbf{p}}(\mathbf{r})=\left(\begin{array}{c} \hat{p}_x(\mathbf{r}) \\ \hat{p}_y(\mathbf{r}) \end{array}\right)=\frac{\hbar}{i}\nabla +\frac{e}{c} \left[\mathbf{A}(\mathbf{r}) +
 \hat{\mathbf{A}}^{\mathrm{ph}}(\mathbf{r})\right],   \label{mom}
\end{equation}
to zero.
The Hamiltonian in \eq{H^S} includes a kinetic part, 
a static external magnetic field $\mathbf{B} = B
\hat{\mb{z}}$, in Landau gauge being represented by the vector potential
$\mathbf{A}(\mathbf{r})= -By \mathbf{e}_x$ and a photon field.
Furthermore, in Eq. (\ref{H^S}), 
\begin{equation}
 H_{Z}=\frac{\mu_B g_S B}{2}\sigma_{z} \label{H_Z}
\end{equation}
describes the Zeeman interaction between the spin and the magnetic field, 
where $g_S$ is the electron spin g-factor and $\mu_B=e\hbar/ (2m_{e}c)$ is the Bohr magneton.
The interaction between the spin and the orbital motion is described by the Rashba part
\begin{equation}
 \hat{H}_{R}(\mathbf{r})=\frac{\alpha}{\hbar}\left( \sigma_{x} \hat{p}_y(\mathbf{r}) -\sigma_{y} \hat{p}_x(\mathbf{r}) \right) \label{H_R}
\end{equation}
with the Rashba coefficient $\alpha$ and Dresselhaus part
\begin{equation}
 \hat{H}_{D}(\mathbf{r})=\frac{\beta}{\hbar}\left(\sigma_{x} \hat{p}_x(\mathbf{r}) - \sigma_{y}  \hat{p}_y(\mathbf{r})\right) \label{H_D}
\end{equation}
with the Dresselhaus coefficient $\beta$. In Eqs.\ (\ref{H_Z}-\ref{H_D}), $\sigma_x$, $\sigma_y$ and $\sigma_z$
represent the spin Pauli matrices.

Equation (\ref{H^S}) includes the electron-electron interaction
\begin{equation}
 \hat{H}_{ee}=\frac{e^2}{2\kappa}\int dx' \; \int dx\; \frac{\hat{\Psi}^{\dagger}(x)\hat{\Psi}^{\dagger}(x')\hat{\Psi}(x')\hat{\Psi}(x)}{\sqrt{|\mathbf{r}-\mathbf{r'}|^2+\eta^2}} \label{Hee}
\end{equation}
with $e>0$ being the magnitude of the electron charge, which is treated numerically exactly. 
Only for numerical reasons, 
we include a small regularization parameter $\eta=0.2387$~nm in \eq{Hee}.
The last term in \eq{H^S} indicates the quantized photon field,
where $\hat{a}^{\dagger}$ is the
photon creation operator
and $\hbar\omega$ is the photon excitation energy. The photon field
interacts with the electron system via the vector potential
\begin{equation}
 \hat{\mathbf{A}}^{\mathrm{ph}}=A(\mathbf{e}\hat{a}+\mathbf{e}^{*}\hat{a}^{\dagger}) \label{vec_pot}
\end{equation}
with
\begin{equation}
 \mathbf{e}=\left\{  \begin{array}{cl} \mathbf{e}_x, & \mathrm{TE}_{011} \\ \mathbf{e}_y, & \mathrm{TE}_{101} \\
             \frac{1}{\sqrt{2}}\left[\mathbf{e}_x+ i\mathbf{e}_y\right], & \textrm{RH circular} \\
             \frac{1}{\sqrt{2}}\left[\mathbf{e}_x- i\mathbf{e}_y\right], & \textrm{LH circular} 
\end{array} \right.
\end{equation}
for a longitudinally-polarized ($x$-polarized) photon
field ($\mathrm{TE}_{011}$), transversely-polarized ($y$-polarized)
photon field ($\mathrm{TE}_{101}$), right-hand (RH) or left-hand (LH) circularly polarized photon field.
The electron-photon coupling
constant $g^{EM}=eA a_w \Omega_w/c$ scales with the amplitude $A$ of
the electromagnetic field. It is interesting to note that
the photon field couples directly to the spin via Eqs.\ (\ref{H_R}), (\ref{H_D}) and (\ref{mom}).
For reasons of comparison and to determine the photocurrents, we also
consider results without photons in the system. In this case,
$\hat{\mathbf{A}}^{\mathrm{ph}}(\mathbf{r})$ and $\hbar \omega
\hat{a}^{\dagger}\hat{a}$ drop out from the MB system Hamiltonian in
\eq{H^S}. Our model of a photon cavity can be realized
experimentally~\cite{0034-4885-69-5-R02, 0953-4075-38-9-007, 0953-8984-20-45-454209}
by letting the photon cavity be much larger than the quantum ring
(this assumption is used in the derivation of the vector potential, \eq{vec_pot}).

\subsection{Quantum ring potential}

\begin{figure}[htbq] %bb= 94 80 508 377
       \includegraphics[width=0.45\textwidth,angle=0,bb= 94 80 508 377]{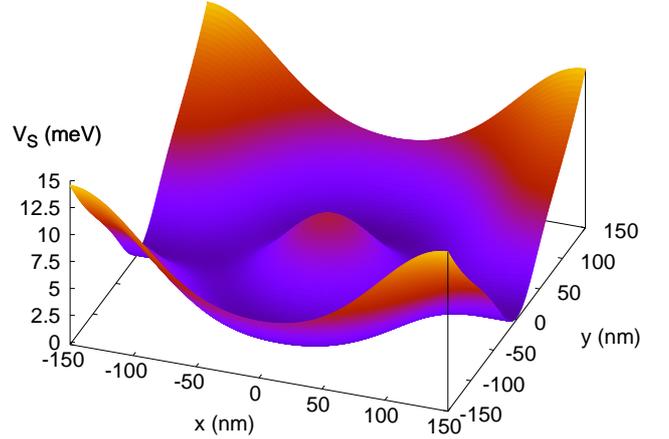}
       \caption{(Color online) System potential $V_S(\mathbf{r})$ of the central ring
       system connected to the left and right leads.}
       \label{ext_pot}
\end{figure}

The quantum ring is embedded in the central system of length $L_x = 300$~\textrm{nm}
situated between two contact areas that will be coupled to the
external leads, as is depicted in \fig{ext_pot}. The system
potential is described by
\begin{eqnarray}
 V_S(\mathbf{r})&=& \sum_{i=1}^{6}V_{i}\exp
 \left[
 -\left(\beta_{xi}(x-x_{0i})\right)^2 - \left(\beta_{yi}y\right)^2
 \right] \nonumber \\
 &&+\frac{1}{2}m^* \Omega_{0}^2y^2, \label{V_S}
\end{eqnarray}
with the parameters from Table \ref{table:ringpot}.
$x_{03}=\epsilon$ is a small numerical symmetry breaking parameter and
$|\epsilon|=10^{-5}$~nm is enough for numerical stability. 
In \eq{V_S}, $\hbar\Omega_0 = 1.0$~meV is the characteristic energy of the confinement
in the system.
\begin{table}
 \caption{Parameters of the ring potential in the central region.}
 \centering
 \begin{tabular}{c    |    c    |    c    |    c    |    c}
\hline\hline
%\cline{1-5}
\\ [-1.4ex]
$i$ &  $V_{i}$ in meV &  $\beta_{xi}$ in $\frac{1}{\mathrm{nm}}$
    &  $x_{0i}$ in nm &  $\beta_{yi}$ in $\frac{1}{\mathrm{nm}}$ \\ [0.5ex]
\hline
1 & 10 & 0.013 & 150 & 0 \\
2 & 10 & 0.013 & -150 & 0 \\
3 & 11.1 & 0.0165 & $\epsilon$ & 0.0165 \\
4 & -4.7 & 0.02 & 149 & 0.02 \\
5 & -4.7 & 0.02 & -149 & 0.02 \\
6 & -5.33 & 0 & 0 & 0 \\[0.5ex]
\hline\hline
\end{tabular}
\label{table:ringpot}
\end{table}

\subsection{Lead Hamiltonian}

The Hamiltonian for the semi-infinite lead $l\in\{L,R\}$ (left or right lead),
\begin{eqnarray}
\hat{H}_{l}&=&\int d^2 r\; \int d^2 r'\; \hat{\mathbf{\Psi}}_{l}^{\dagger}(\mathbf{r}')\delta(\mathbf{r}'-\mathbf{r})\left[\left(\frac{\hat{\mathbf{p}}_l^2}{2m^{*}} +V_l(\mathbf{r})\right)\nonumber \right. \\
&&+\left.H_{Z}+\hat{H}_{R}(\mathbf{r})+ \hat{H}_{D}(\mathbf{r})\right]\hat{\mathbf{\Psi}}_{l}(\mathbf{r}), \label{H^l}
\end{eqnarray}
with the momentum operator containing the kinetic momentum 
and the vector potential leading only to the magnetic field (i.e.\ not to the photon field)
\begin{equation}
 \hat{\mathbf{p}}_l(\mathbf{r})=\frac{\hbar}{i}\nabla +\frac{e}{c} \mathbf{A}(\mathbf{r}).   \label{mom_l}
\end{equation}
We remind the reader that the Rashba part, $\hat{H}_{R}(\mathbf{r})$, (\eq{H_R}) 
and Dresselhaus part, $\hat{H}_{D}(\mathbf{r})$, (\eq{H_D})
of the SOI are momentum-dependent. For the leads, the momentum from \eq{mom_l},
is used for the Rashba and Dresselhaus terms in \eq{H^l}. Equation (\ref{H^l}) contains the lead field operator
\begin{equation}
 \hat{\Psi}_{l}(x)=\sum_{q}\psi_{ql}(x)\hat{C}_{ql} \label{FOl}
\end{equation}
in the spinor
\begin{equation}
 \hat{\mathbf{\Psi}}_{l}(\mathbf{r})=\left(\begin{array}{c} \hat{\Psi}_{l}(\uparrow,\mathbf{r}) \\ \hat{\Psi}_{l}(\downarrow,\mathbf{r}) \end{array}\right) \label{FOS_leads}
\end{equation}
and a corresponding definition to \eq{conj_FOS} for the Hermitian conjugate of $\hat{\mathbf{\Psi}}_{l}(\mathbf{r})$ in \eq{FOS_leads}.
In \eq{FOl}, $\psi_{ql}(x)$ is a SES in the lead $l$ (eigenstate with quantum number $q$ of the Hamiltonian of \eq{H^l})
and $\hat{C}_{ql}$ is the associated electron annihilation operator.
The lead potential
\begin{equation}
 V_{l}(\mathbf{r})=\frac{1}{2} m^{*} \Omega_{l}^2 y^2
\end{equation}
confines the electrons parabolically in $y$-direction in the leads
with the characteristic energy $\hbar\Omega_l = 2.0$~meV.

\subsection{Time-convolutionless generalized master equation approach}

We use the time-convolutionless generalized master equation~\cite{PhysRevA.59.1633} (TCL-GME),
which is a non-Markovian master equation that is local in time.
This master equation satisfies the positivity conditions~\cite{Whitney08:175304}
for the MB state occupation probabilities in the reduced density operator (RDO) usually
to a relatively strong system-lead coupling~\cite{PhysRevB.87.035314}.
We assume the
initial total statistical density matrix to be 
a product state of the system and leads density matrices, before
we switch on the coupling to the leads,
\begin{equation}
 \hat{W}(0)=\hat{\rho}_L \otimes \hat{\rho}_R \otimes \hat{\rho}_S(0),
\end{equation}
with $\rho_l$, $l\in \{L,R\}$, being the normalized density matrices
of the leads. The coupling Hamiltonian between the central system
and the leads reads
\begin{equation}
 \hat{H}_{T}(t)= \sum_{l=L,R}\int dq\;\chi^{l}(t)\left[\hat{\mathfrak{T}}^{l}(q)\hat{C}_{ql}
 +\hat{C}_{ql}^{\dagger} \hat{\mathfrak{T}}^{l\dagger}(q) \right]\,
 .
\end{equation}
The coupling is switched on at $t=0$ via the switching function
\begin{equation}
 \chi^{l}(t)=1-\frac{2}{e^{\alpha^{l}t}+1}
\end{equation}
with a switching parameter $\alpha^l$ and
\begin{equation}
 \hat{\mathfrak{T}}^{l}(q)
 = \sum_{\alpha \beta}\ketmes{\alpha}\brames{\beta}\sum_{a}T_{qa}^{l}
 \brames{\alpha} \hat{C}_{a}^{\dagger}\ketmes{\beta}.
 \label{mathfrakT}
\end{equation}
Equation (\ref{mathfrakT}) is written in
the MB eigenbasis $\{\ketmes{\alpha}\}$ of the system Hamiltonian, \eq{H^S}.  The
coupling tensor~\cite{Gudmundsson85:075306}
\begin{eqnarray}
 T_{qa}^{l}&=&\sum_{\sigma}\sum_{\sigma'}\int_{\Omega^l} d^2r\; \int_{\Omega_S^l} d^2r'\;
 \psi_{ql}^{*}(\mathbf{r},\sigma) \nonumber \\
 && \times g_{aq}^{l}(\mathbf{r},\mathbf{r'},\sigma,\sigma')\psi_a^S(\mathbf{r}',\sigma')
\end{eqnarray}
couples the lead SES
$\{\psi_{ql}(\mathbf{r},\sigma)\}$ with energy spectrum $\{\epsilon^l(q)\}$
to the system SES $\{\psi_a^S(\mathbf{r},\sigma)\}$ with energy spectrum
$\{E_a\}$ that reach into the contact regions~\cite{1367-2630-11-11-113007}, $\Omega_S^l$ and $\Omega_l$, of system
and lead $l$, respectively, and the coupling kernel
\begin{eqnarray}
 g_{aq}^l(\mathbf{r},\mathbf{r'},\sigma,\sigma')&=&g_0^l\delta_{\sigma,\sigma'}\exp\left[-\delta_x^l(x-x')^2\right]\nonumber \\
&&\times \exp\left[-\delta_y^l(y-y')^2 \right]\nonumber \\
&&\times \exp{
\left(
 -\frac{|E_a-\epsilon^l(q)|}{\Delta^l_E}
\right)} \label{gaql}
\end{eqnarray}
suppresses different-spin coupling. Note that the meaning of $x$ in \eq{gaql}
is $\mathbf{r}=(x,y)$ and not $x=(\mathbf{r},\sigma)$.
In \eq{gaql}, $g_0^l$ is the lead coupling strength and $\delta^l_x$
and $\delta^l_y$ are the contact region parameters for lead $l$ in $x$-
and $y$-direction, respectively. Moreover, $\Delta^l_E$ denotes the
affinity constant between the central system SES energy levels $\{E_a\}$
and the lead energy levels $\{\epsilon^l(q)\}$.

When propagated with the TCL-GME~\cite{PhysRevA.59.1633,PhysRevB.87.035314},
the RDO of the system,
\begin{equation}
 \hat{\rho}_S(t)= \mathrm{Tr}_{L}\mathrm{Tr}_{R}[\hat{W}(t)],
\end{equation}
evolves to second order in the lead coupling strength via
\begin{eqnarray}
\dot{\hat{\rho}}_{S}(t)&=&-\frac{i}{\hbar}[\hat{H}_{S},\hat{\rho}_{S}(t)]\nonumber \\
&&-\Bigg[\sum_{l=L,R} \int dq\;\Big[\hat{\mathfrak{T}}^{l}(q),
\hat{\Omega}^{l}(q,t)\hat{\rho}_{S}(t) \nonumber \\ &&-
f(\epsilon^{l}(q)) \left\{ \hat{\rho}_{S}(t),\hat{\Omega}^{l}(q,t)
\right\} \Big] +\mathrm{H.c.}\Bigg]
\end{eqnarray}
with the Fermi distribution function $f(E)$,
\begin{eqnarray}
 \hat{\Omega}^{l}(q,t) &=& \frac{1}{\hbar^2}\chi^{l}(t)\exp\left(-\frac{i}{\hbar}t\epsilon^{l}(q)\right) \nonumber \\
 && \times \hat{U}_{S}(t) \hat{\Pi}^{l}(q,t)  \hat{U}_{S}^{\dagger}(t)
\end{eqnarray}
and
\begin{eqnarray}
 \hat{\Pi}^{l}(q,t)&=& \int_{0}^{t}dt'\; \left[ \exp\left(\frac{i}{\hbar}t'\epsilon^{l}(q)\right) \chi^{l}(t') \right. \nonumber \\
 && \times \left. \hat{U}_{S}^{\dagger}(t') \hat{\mathfrak{T}}^{l\dagger}(q) \hat{U}_{S}(t') \right].
\end{eqnarray}

\subsection{Transport quantities used in the numerical results}

We investigate numerically the non-equilibrium electron
transport properties through a quantum ring system, 
which is situated in a photon
cavity and weakly coupled to leads for the parameters given in the
\ref{parnr}.
To explore the influence of the circularly polarized photon field and
Rashba and Dresselhaus coupling on the non-local charge and spin polarization
current from and into the leads, 
we define the non-local right-going charge current $I_{l}^{c}(t)$ (lead-ring charge current)
in lead $l=L,R$ by
\begin{equation}
 I_{l}^{c}(t)=c_{l}\mathrm{Tr}[\dot{\hat{\rho}}_S^l(t) \hat{Q}]
\end{equation}
with $c_L=1$ and $c_R=-1$, with the charge operator
\begin{equation}
 \hat{Q}=\int d^2r\;  \hat{n}^{c}(\mathbf{r}) \label{chcs}
\end{equation}
and the time-derivative of the RDO in the MB
basis due to the coupling to the lead $l\in\{L,R\}$
\begin{eqnarray}
 \dot{\hat{\rho}}_S^l(t)&=&-\int dq\;\Big[\mathfrak{T}^{l}(q), \Big[
\Omega^{l}(q,t)\rho_{S}(t) - \nonumber \\
 && f(\epsilon^{l}(q)) \left\{ \rho_{S}(t),\Omega^{l}(q,t) \right\}
\Big] \Big]+\mathrm{H.c.}.
\end{eqnarray}
The charge density operator $\hat{n}^{c}(\mathbf{r})$ in \eq{chcs} is given in the \ref{chspdcdop}, \eq{ncq}.
Similarly, we define the non-local right-going spin polarization current $I_{l}^{i}(t)$ for $S_i$ spin polarization
(lead-ring spin polarization current) in lead $l=L,R$ by
\begin{equation}
 I_{l}^{i}(t)=c_{l}\mathrm{Tr}[\dot{\hat{\rho}}_S^l(t) \hat{S}_{i}] \label{nlspincur}
\end{equation}
with $i=x,y,z$ and the spin polarization operator for $S_i$ spin polarization
\begin{equation}
 \hat{S}_{i}=\int d^2r\; \hat{n}^{i}(\mathbf{r}),
\end{equation}
where the spin polarization density operator for spin polarization $S_i$, $\hat{n}^{i}(\mathbf{r})$, is defined in \eq{niq} in the \ref{chspdcdop}. 
To get more insight into the local current flow in the ring system,
we define the top local charge ($\gamma=c$) and spin
($\gamma=x,y,z$, where $\gamma$ describes the spin polarization) 
current through the upper arm ($y>0$) of the ring
\begin{equation}
 I_{\mathrm{top}}^{\gamma}(t)=\int_{0}^{\infty}dy\;j_x^{\gamma}(x=0,y,t)
\end{equation}
and the bottom local charge and spin polarization current through the lower arm ($y<0$) of
the ring
\begin{equation}
 I_{\mathrm{bottom}}^{\gamma}(t)=\int_{-\infty}^{0}dy\;j_x^{\gamma}(x=0,y,t)\, .
\end{equation}
Here, the charge and spin polarization current density,
\begin{equation}
 \mathbf{j}^{\gamma}(\mathbf{r},t)
 =\left(\begin{array}{c} j_x^{\gamma}(\mathbf{r},t)\\j_y^{\gamma}(\mathbf{r},t) \end{array}\right)
 =\mathrm{Tr}[\hat{\rho}_{S}(t)\hat{\mathbf{j}}^{\gamma}(\mathbf{r})],
\end{equation}
is given by the expectation value of the charge and spin polarization current density
operator, \eq{jcq}, \eq{jsxq}, \eq{jsyq} and \eq{jszq} in the \ref{chspdcdop}.
We note that while the charge density is satisfying the continuity equation
\begin{equation}
 \frac{\partial}{\partial t} n^{c}(\mathbf{r},t)+\nabla \mathbf{j}^{c}(\mathbf{r},t)=0, \label{chargecont}
\end{equation}
the continuity equation for the spin polarization density includes in general the source terms
\begin{equation}
 s^{i}(\mathbf{r},t)=\frac{\partial}{\partial t} n^{i}(\mathbf{r},t)+\nabla \mathbf{j}^{i}(\mathbf{r},t). \label{spincont}
\end{equation}
The definition for the spin polarization current density (\eq{jsxq}, \eq{jsyq} and \eq{jszq} from the appendix)
corresponds to the minimal (simplest) expression for the source operator~\cite{2013arXiv1310.5870A}
and agrees with the definition of the Rashba current when we limit ourselves to the case without
magnetic and photon field and without Dresselhaus SOI~\cite{PhysRevB.68.241315, PhysRevB.76.033306}.
Furthermore, to distinguish better the structure
of the dynamical transport features,
it is convenient to define the total local (TL) charge or spin polarization current
\begin{equation}
 I_{\rm tl}^{\gamma}(t)=I_{\rm top}^{\gamma}(t)+I_{\rm bottom}^{\gamma}(t)
\end{equation}
and circular local (CL) charge or spin polarization current
\begin{equation}
 I_{\rm cl}^{\gamma}(t)=\frac{1}{2}\left[I_{\rm bottom}^{\gamma}(t)-I_{\rm top}^{\gamma}(t)\right],
\end{equation}
which is positive if the electrons move counter-clockwise in the ring.
The TL charge current is usually bias driven while the CL charge current
can be driven by the circularly polarized photon field
(or a strong magnetic field).
The TL spin polarization current is usually related to non-vanishing spin polarization sources
while a CL spin polarization current can exist without such sources.

To explore the influence of the photon field, we define
the TL charge or spin photocurrent
\begin{equation}
 I_{\rm ph, tl}^{\gamma,p}(t)=I_{\rm tl}^{\gamma,p}(t)-I_{\rm tl}^{\gamma,0}(t)
\end{equation}
and CL charge or spin photocurrent
\begin{equation}
 I_{\rm ph, cl}^{\gamma,p}(t)=I_{\rm cl}^{\gamma,p}(t)-I_{\rm cl}^{\gamma,0}(t),
\end{equation}
which are given by the difference of the associated local currents 
with ($I^{\gamma,p}(t)$) and without ($I^{\gamma,0}(t)$) photons, where
$p=x,y,r,l$ denotes the polarization of the photon field
($x$: $x$-polarization, $y$: $y$-polarization, $r$: RH circular polarization,
$l$: LH circular polarization).
The total charge of the central system is given by
\begin{equation}
 Q(t)=\mathrm{Tr}[\hat{\rho}_S(t) \hat{Q}]
\end{equation}
and the spin polarization of the central system
\begin{equation}
 S_{i}(t)=\mathrm{Tr}[\hat{\rho}_S(t) \hat{S}_{i}].
\end{equation}

\section{Spin polarization} \label{sec3}

In this section, we show the spin polarization of the central ring system for linearly or circularly
polarized photon field as a function of the Rashba or Dresselhaus parameter. 
The ring is connected to leads, in which a chemical potential bias is maintained. 
The spin polarization $\mathbf{S}=(S_x, S_y, S_z)$ is a three-dimensional vector,
which, in the Rashba case, is
influenced by the effective magnetic field associated with the Rashba effect.

\subsection{Linear photon field polarization}

\begin{figure}[htbq] %bb=1 71 675 355,clip
       \includegraphics[width=0.34\textwidth,angle=-90]{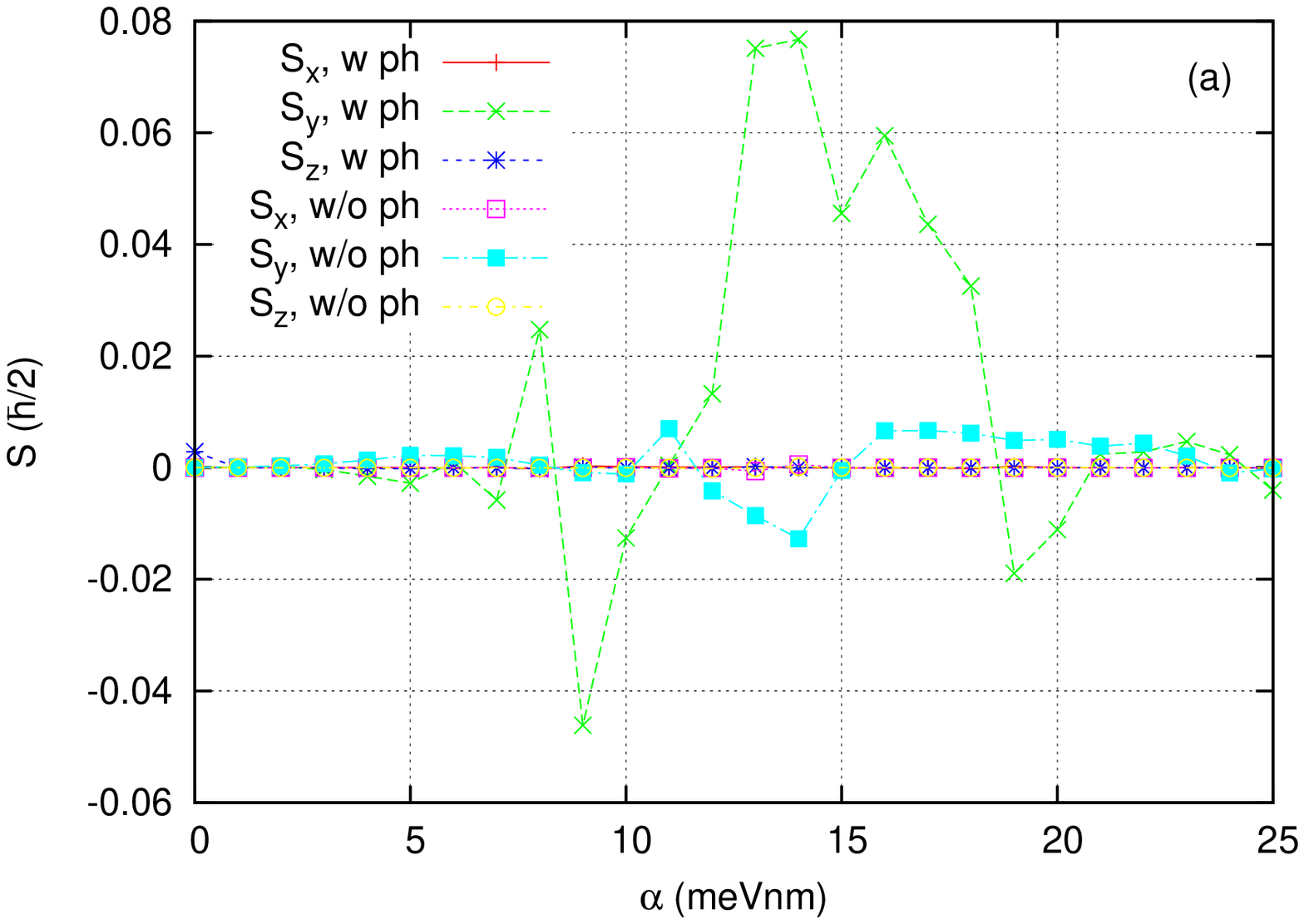}
       \includegraphics[width=0.34\textwidth,angle=-90]{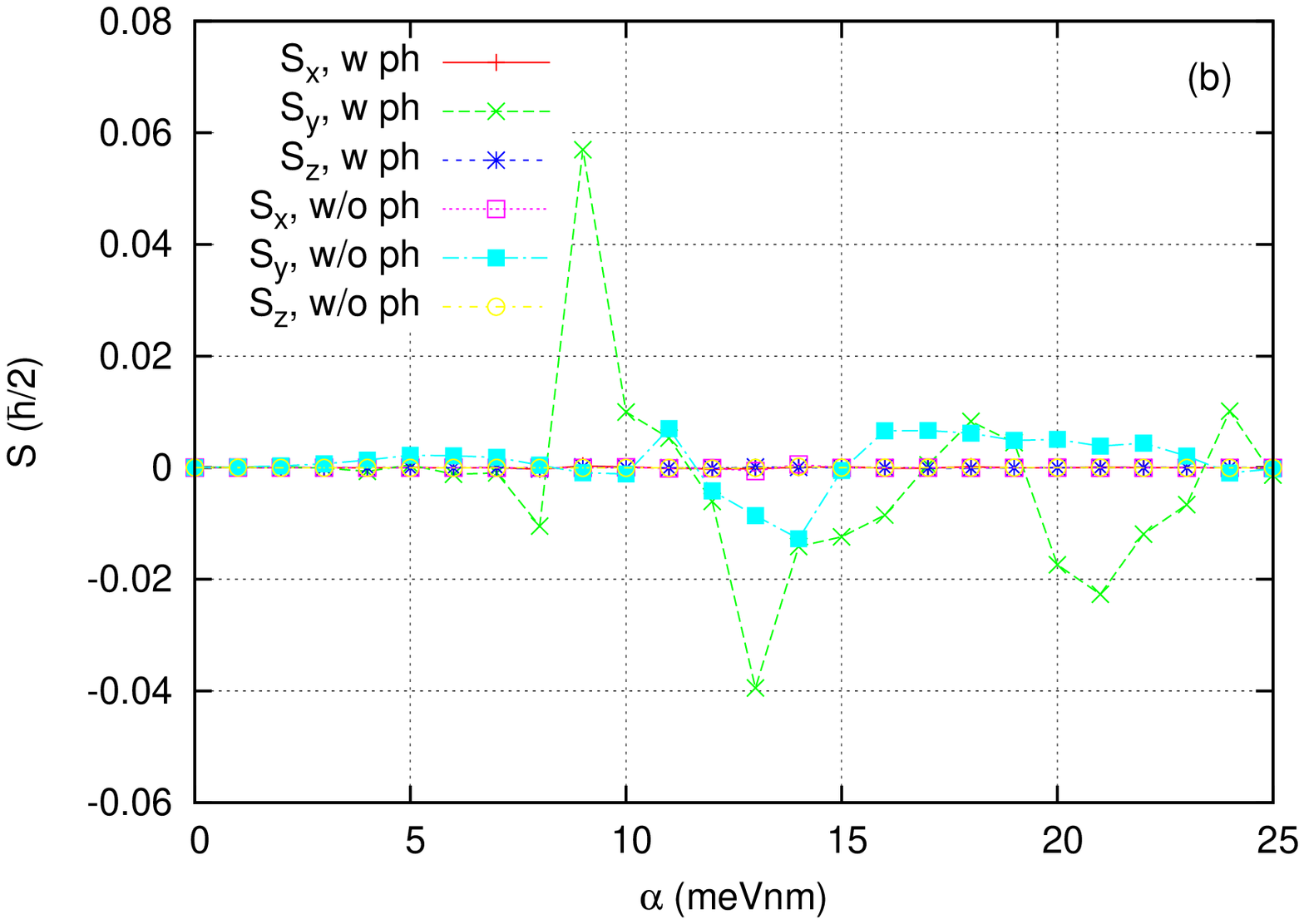}
       \caption{(Color online) Spin polarization $\mathbf{S}=(S_x, S_y, S_z)$ of the central system 
       versus the Rashba coefficient $\alpha$
       averaged over the time interval $[180,220]$~ps
       with (w) (a) $x$-polarized photon field 
       and (b) $y$-polarized photon field
       or without (w/o) photon cavity
       ($\beta=0$ and $B=10^{-5}$~T).}
       \label{den_tot_alpha_200}
\end{figure}

Here, we will compare the spin polarization in the central system 
for $x$- or $y$-polarization of the photon field with the spin polarization
in the case that the photon cavity is removed. 
Figure \ref{den_tot_alpha_200} shows the spin polarization
as a function of the Rashba coefficient.
The critical value of the Rashba coefficient, which describes
the position of the destructive AC interference is $\alpha^c \approx  13$~meV\,nm, 
where the TL charge current has a pronounced minimum~\cite{2013arXiv1310.5870A}.
The spin polarization is largest around $\alpha=\alpha^c$
due to spin accumulation in the current suppressed regime. 
For $\alpha \to 0$, the spin polarization should vanish, except for the minor spin polarization in $S_z$ due to the Zeeman interaction with the small magnetic field.
Apart from that, only the $S_y$ spin polarization seems to be significant in \fig{den_tot_alpha_200}.

The direction of the spin polarization vector can be explained with the concept of the
effective magnetic field,
\begin{equation}
 \hat{\mathbf{B}}_{R}=-\frac{\hat{\mathbf{p}}\times \mathbf{E}}{m^* c},
\end{equation}
occurring due to the electronic motion in the electric field $\mathbf{E}=E \mathbf{e}_z$. Consequently, we can write the Rashba term
\begin{equation}
 \hat{H}_R=\frac{\alpha'}{\hbar}\mathbf{\boldsymbol{\sigma}} \cdot (\hat{\mathbf{p}} \times \mathbf{E})=-\frac{\alpha' m^{*}c}{\hbar}\boldsymbol{\sigma} \cdot \hat{\mathbf{B}}_{R}
\end{equation}
with $\alpha'=\alpha/E$. The spin polarization densities vanish 
in a one-dimensional (1D) ring with only Rashba SOI
due to Kramers degeneracy for the time-reversal symmetric system
(see also Ref.~\cite{2013arXiv1310.5870A}).
It is therefore clear that the spin polarization densities
in the case without photon cavity result only from 
the geometric deviations from the 1D ring model, for example, the contact regions.
The main transport and canonical momentum in the contact regions is along the $x$-direction. 
As a consequence, the Rashba effective magnetic field, $\hat{\mathbf{B}}_{R}$, should be parallel to the $y$-direction
and induce a spin polarization in mainly the $y$-direction as is in fact depicted in \fig{den_tot_alpha_200}. 
With photon cavity, the $x$-polarized photons
should lead to an additional kinetic momentum of the electrons in $x$-direction
increasing the $S_y$ spin polarization further. This is also very well in agreement with \fig{den_tot_alpha_200}(a). However, it is interesting that the $y$-polarized photons do
\textit{not} induce an $S_x$ spin polarization although the vector potential contribution to the kinetic momentum
would suggest this. 
Here, the reason for the vanishing $S_x$ and $S_z$ spin polarization is that the
spin polarization density distribution for the $S_x$ and $S_z$ spin polarization is 
constrained to an antisymmetric function in $y$
around the x-axis ($y=0$) for any time $t$, when the central system is initially empty ($N_{e,\rm{init}}=0$). 
The spin polarization density distribution for spin polarization $S_y$ is symmetric around $y=0$ 
permitting the non-vanishing spin-polarization $S_y$.
As a result, the $y$-polarized photon field increases only the $S_y$ spin polarization, but
less than in the $x$-polarized case (\fig{den_tot_alpha_200}(b).
The symmetry properties and, as a consequence, the non-vanishing components of the spin polarization change if $N_{e,\rm{init}}>0$. 
Alternatively, $S_x\not\approx 0$ and $S_z\not\approx 0$ could be achieved with a circularly polarized photon field.

\subsection{Circular photon field polarization}

\begin{figure}[htbq]
       \begin{center}
       \includegraphics[width=0.245\textwidth,angle=-90,bb=208 57 552 538,clip]{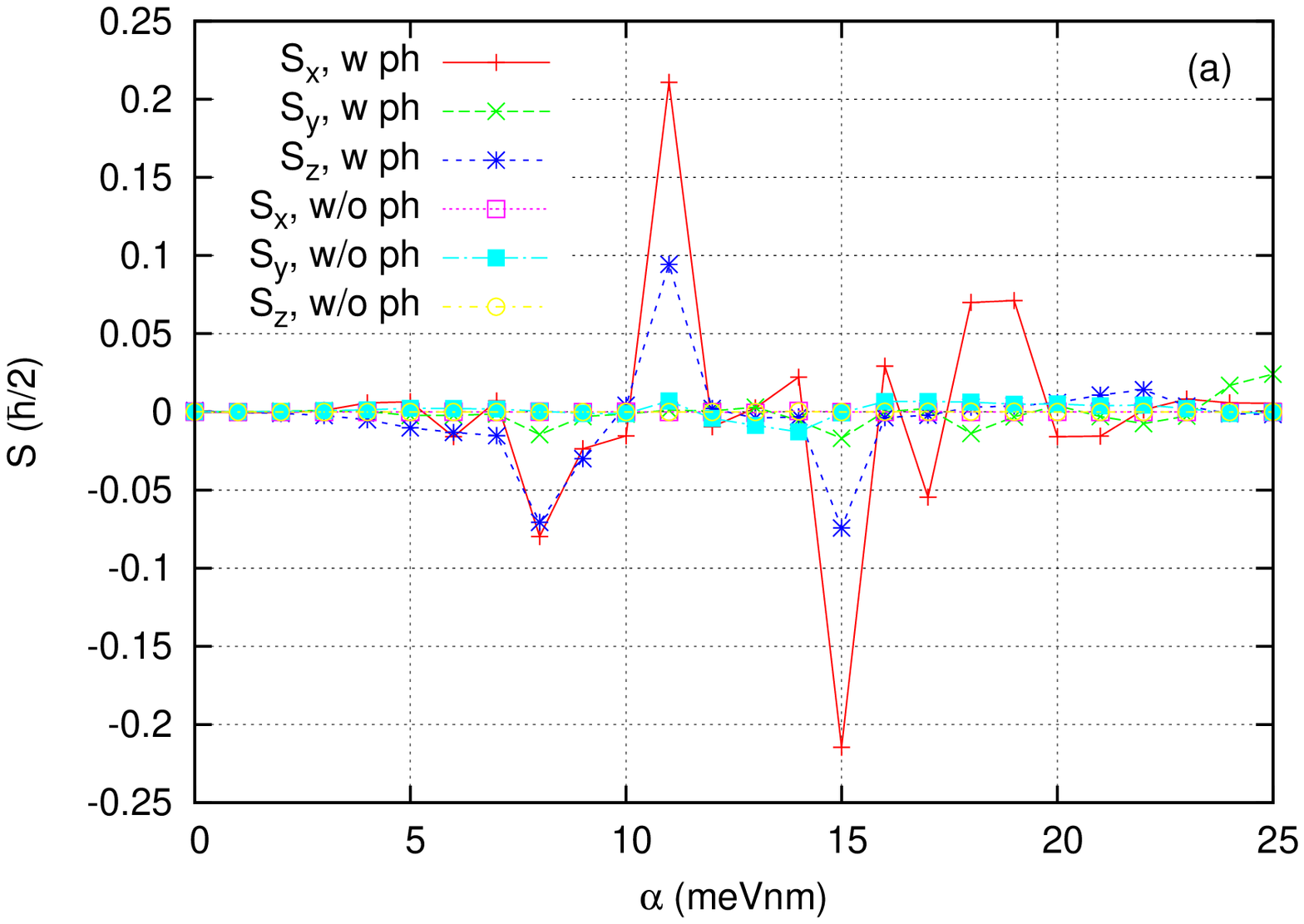}%var width=0.35\textwidth
       
       \includegraphics[width=0.245\textwidth,angle=-90,bb=208 57 552 538,clip]{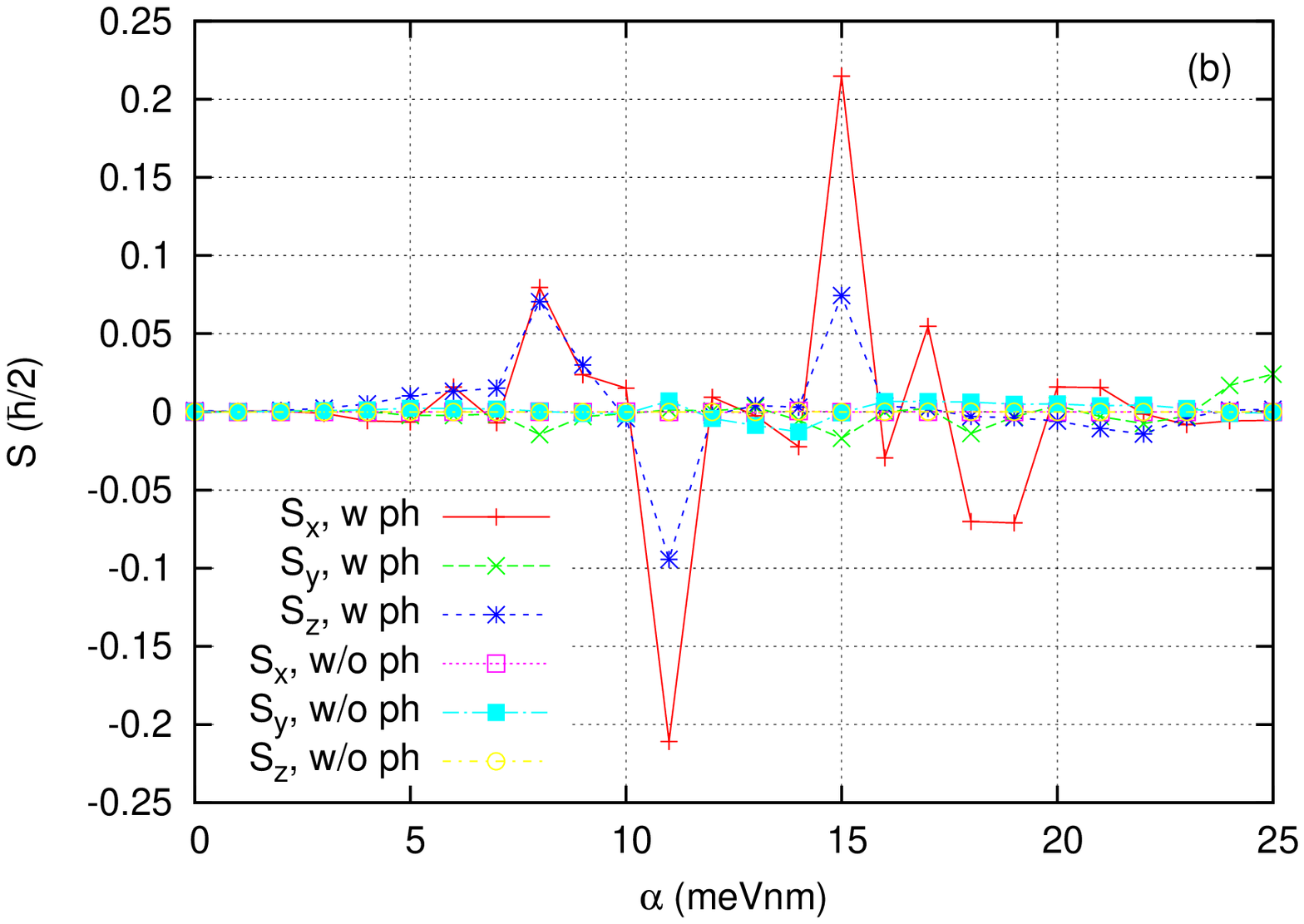}%var width=0.35\textwidth
       
       \includegraphics[width=0.245\textwidth,angle=-90,bb=208 57 552 538,clip]{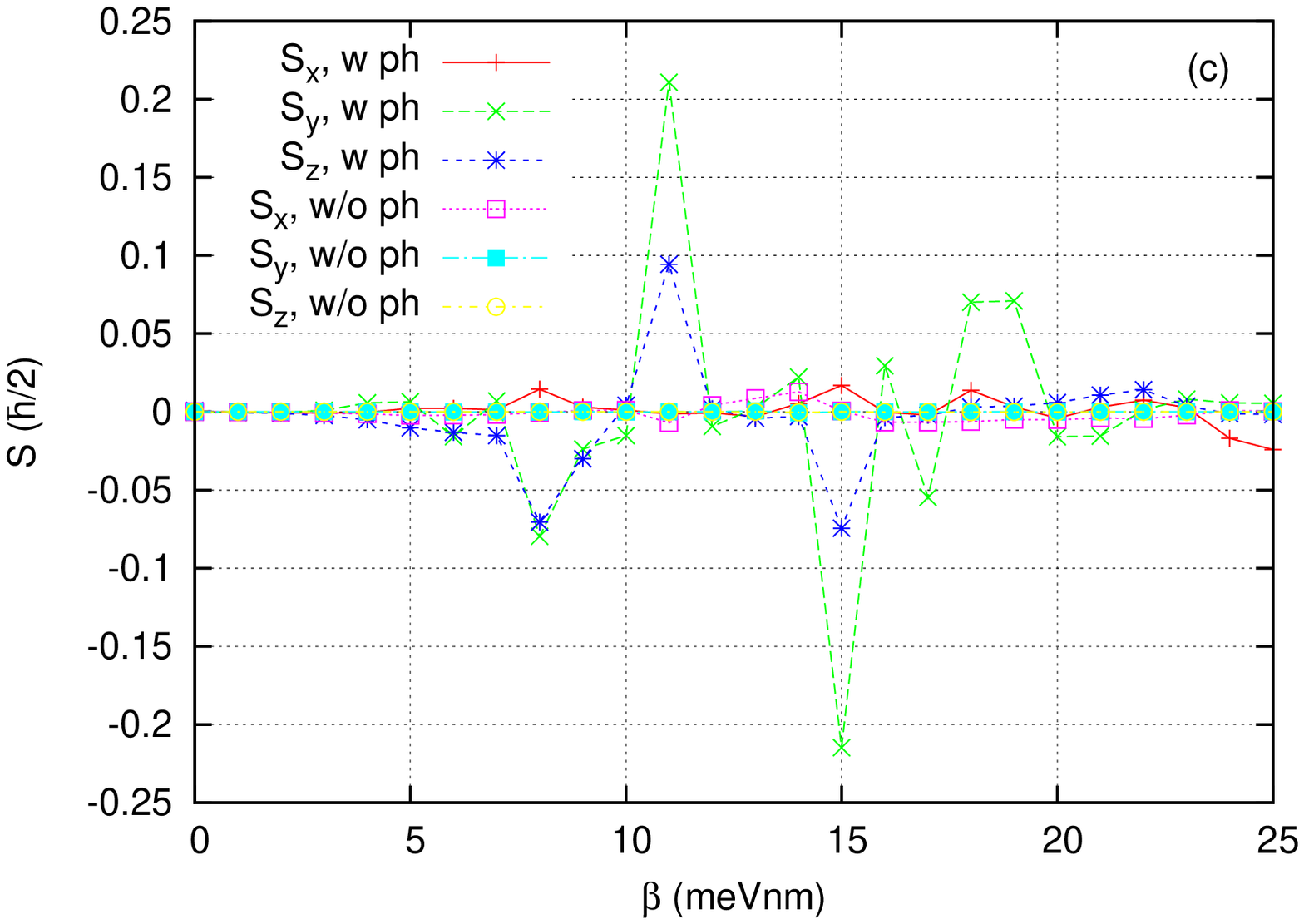}%var width=0.35\textwidth
       \end{center}
       \caption{(Color online) Spin polarization $\mathbf{S}=(S_x, S_y, S_z)$ of the central system
       averaged over the time interval $[180,220]$~ps 
       with (w) or without (w/o) photon cavity
       and (a) LH circularly polarized photon field versus the
       Rashba coefficient $\alpha$ ($\beta=0$),  
       (b) RH circularly polarized photon field versus the
       Rashba coefficient $\alpha$ ($\beta=0$) and
       (c) RH circularly polarized photon field versus the
       Dresselhaus coefficient $\beta$ ($\alpha=0$).}
       \label{den_tot_c_200}
\end{figure}

The circularly polarized photon field has a non-vanishing spin angular momentum perturbing the
angular orbital motion of the electrons. Electrons in a 1D ring geometry
(as an approximation of our geometry) do not show a circular charge current for
vanishing magnetic field~\cite{2013arXiv1310.5870A}, 
but placing them in a photon cavity with circularly polarized photon field
would let them move around the ring due to the spin angular momentum of light.
The circular motion is a much stronger perturbation of the ring electrons 
than the perturbation caused by the linearly polarized photon field.
The angular electronic motion in the electric field causes an effective magnetic field
and a local spin polarization in radial direction.
In our 2D geometry, we will see that the circularly polarized photon field induces also
vortices of the size of the ring width. As a consequence, the spin polarization is
not only a local quantity. 
It should be substantially larger than for linear polarization due to the 
strong perturbation of the circularly polarized photon field with the electronic system.

\begin{figure}[htbq]
       \begin{center}%bb=89 59 252 219
       \includegraphics[width=0.57\textwidth,angle=0]{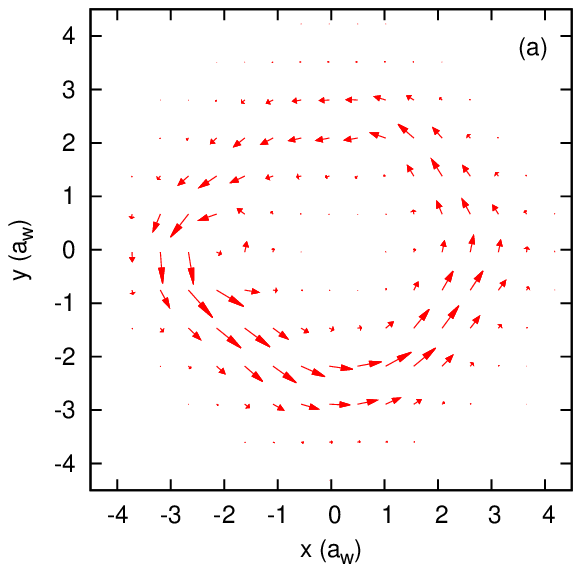}
       \includegraphics[width=0.57\textwidth,angle=0]{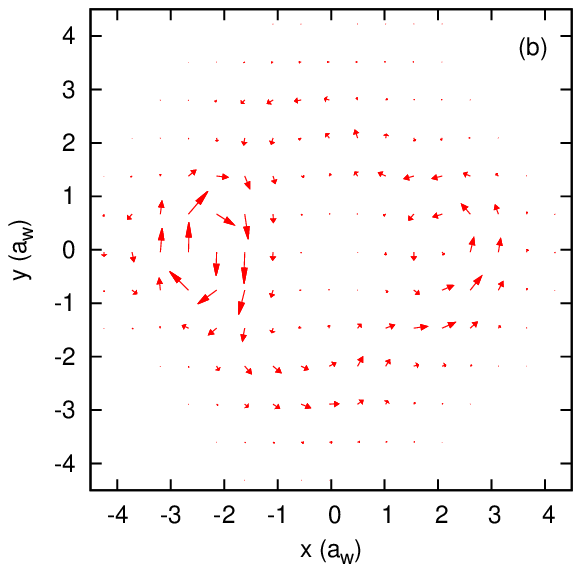}
       \end{center}
       \caption{(Color online) Normalized vector fields of the 
       charge current density $\mathbf{j}^{c}(x,y)$
       in the central system for (a) the Rashba coefficient
       $\alpha=11$~meV\,nm and (b) $\alpha=15$~meV\,nm
       at $t=200$~ps 
       with RH circularly polarized photon field and Dresselhaus coefficient $\beta=0$.
       }
       \label{cur_den_alpha_beta0_B000001p_200_lowres}
\end{figure}

Figure \ref{den_tot_c_200} shows the spin polarization for (a) 
LH circularly polarized photon field and (b) 
and (c) RH circularly polarized photon field.
Different from all other subfigures of \fig{den_tot_alpha_200} and \fig{den_tot_c_200},
\fig{den_tot_c_200}(c) shows the spin polarization as a function of the
Dresselhaus instead of the Rashba coefficient.
The spin polarization is indeed substantially larger for circular polarization
than for linear polarization, \fig{den_tot_alpha_200}. Furthermore, 
the spin polarization could take any direction due to spin precession. This is a fundamental difference between the
linearly and circularly polarized photon field, where only the $S_y$ spin polarization was substantially different from zero.
Only the circularly polarized photon field can induce $S_x$ spin polarization
although the $y$-polarized photon field leads to an effective magnetic field in $x$-direction.

It might seem surprising that the $S_y$ spin polarization is relatively small for the
circularly polarized photon field. 
Figure \ref{cur_den_alpha_beta0_B000001p_200_lowres} shows
the normalized vector field for the charge current density $\mathbf{j}^{c}(x,y)$
for RH circularly polarized photon field and two values of the Rashba coefficient,
symmetrically located around $\alpha = \alpha^c$.
The vortices disappear or are much weaker without photon cavity
or with linearly polarized photon field. They are also weaker for circularly
polarized photon field and Rashba coefficient values, 
which are associated with a smaller spin polarization.
The charge current density is in general a complicated superposition
of many vortices.
We would like to mention that it is important that we have used a ring geometry with a
finite width. 
Otherwise, our numerical calculations would not give a realistic picture of the spin polarization.
The relatively strong vortices in \fig{cur_den_alpha_beta0_B000001p_200_lowres},
which are located close to the contact regions to the leads,
are usually not as symmetric in $x$-direction around their center
than they are in $y$-direction. Correspondingly, there is often a 
relatively strong net $y$-component of the canonical momentum
leading to a Rashba effective magnetic field in $x$-direction and 
a much stronger $S_x$ spin polarization than $S_y$ spin polarization.

It is in particular interesting
to observe the local antisymmetric behavior of the $x$- and $z$-component of the spin polarization 
around the level crossings at $\alpha=\alpha^c$ (\fig{den_tot_c_200}(b)),
which are the spin polarizations induced by the circularly polarized photon field. 
By contrast, the $S_y$ spin polarization is clearly not antisymmetric around $\alpha=\alpha^c$ for the linearly polarized photon field (see \fig{den_tot_alpha_200}). 
It can be seen from a comparison of \fig{cur_den_alpha_beta0_B000001p_200_lowres}(a),
where $\alpha< \alpha^c$,
and \fig{cur_den_alpha_beta0_B000001p_200_lowres}(b),
where $\alpha> \alpha^c$,
that the circulation direction
of the strong vortex at the left contact region is inversed.
This leads to the local
antisymmetric behavior of the $S_x$ and $S_z$ spin polarization around $\alpha= \alpha^c$.
Since the pronounced vortex structure is mainly 
due to the circular polarization of the photon field, only the components of the spin polarization,
which are induced by the circularly polarized cavity photon field
show the local antisymmetric behavior.
The fact that the $S_y$ spin polarization is not antisymmetric for linear
photon field polarization can be understood as follows:
qualitatively, as said before,
only the canonical momentum due to the deviations from the 1D ring geometry 
(and the photon cavity, which is unimportant for $S_y$ in the case of circular photon field polarization)
allows for spin polarization
in the central system.
The contact regions in $x$-direction cause only 
smaller perturbations of the central system spectrum beyond the level-crossing structure from the 1D ring geometry 
(which only the circularly polarized photon field can perturb).
It is not likely that these smaller perturbations would lead to additional
level-crossings around $\alpha\approx 13$~meV\,nm, which we have found to be
responsible for the
antisymmetric behavior of $S_x$ and $S_z$
for circular polarization. 
Therefore, a local antisymmetry of $S_y$ around $\alpha \approx 13$~meV\,nm cannot be found.

The $x$- and $z$-components of the spin polarization
are also antisymmetric with respect to the handedness of the circularly polarized light 
(\fig{den_tot_c_200}(a) in comparison with 
\fig{den_tot_c_200}(b)).
As the $S_x$ and $S_z$ spin polarization are a 
direct and pure consequence of circular polarization meaning
that they are vanishing in the case without photon cavity 
and in the case with linear polarized photon field,
it is understandable that a sign change in the handedness would follow a sign change in the effective
magnetic field and spin polarization. 
In fact, also the vortex circulation direction is inversed by an inversion of the
handedness of the circularly polarized photon field.
The situation is different for the $S_y$ spin polarization, 
which is different from zero in the absence of photons. 
The circularly polarized photon field changes the $S_y$ spin polarization only slightly.
Furthermore, comparing the Rashba and Dresselhaus case for RH circularly polarized photon field (\fig{den_tot_c_200}(b) and \fig{den_tot_c_200}(c)),
we can verify the spin polarization symmetries 
\begin{equation}
\left(\begin{array}{c} S_{xD} \\ S_{yD} \\ S_{zD} \end{array}\right)= -\left(\begin{array}{c} S_{yR} \\ S_{xR} \\ S_{zR} \end{array}\right)
\end{equation}
due to the structure of the Rashba and Dresselhaus Hamiltonian.

\section{Charge and spin polarization currents} \label{sec4}

Here, we show our numerical results for the charge and spin polarization currents,
both between the leads and the system and in the system (quantum ring) itself. 
Emphasis is laid on the phenomena caused by the photon cavity
with a focus on the circularly polarized photon field.

\subsection{Non-local currents}

\begin{figure}[htbq] %bb=1 71 675 355,clip       
       \includegraphics[width=0.34\textwidth,angle=-90]{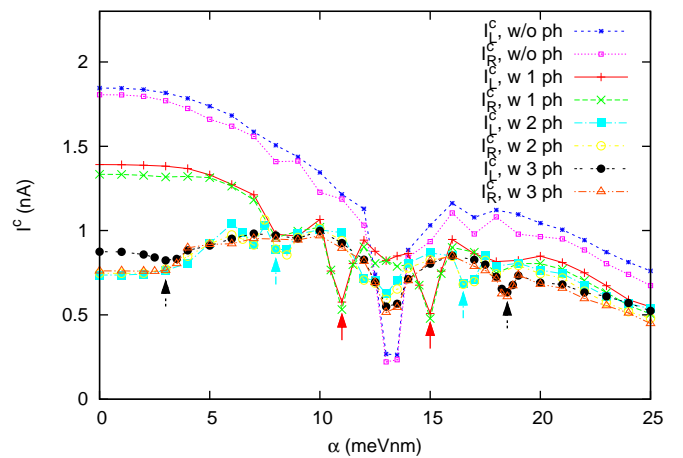}
       \caption{(Color online) Non-local lead-system charge currents $I^c_L$ and
       $I^c_R$
       versus the Rashba coefficient $\alpha$
       at time $t=200$~ps and with RH circularly polarized photon field
       for different initial number of photons $N_{ph,\rm init}=0,1,2,3$.
       The Dresselhaus coefficient $\beta=0$.}
       \label{current_per_200_p_alpha_nph}
\end{figure}

Figure \ref{current_per_200_p_alpha_nph} shows the non-local (lead-ring) charge currents from the
left lead into the system and further into the right lead as a function of the Rashba coefficient $\alpha$
at time $t=200$~ps and circularly polarized photon field. We note in passing that the result
depends not on the handedness of the circularly polarized light.
Left- or right-handedness would only interchange the meaning of the otherwise
indifferent upper and lower ring arm.
With increasing initial number of photons the current tends to get suppressed due to the
in general smaller energy differences between the MB levels and the fact that the states having a larger
photon content lie in general higher in energy.
However, more interestingly, we observe two additional current dips for smaller and larger $\alpha$,
while the current dip at $\alpha=\alpha^c$ appears weaker.
With increasing number of initial photons, the two dips for smaller or larger Rashba coefficient
move to even smaller ($\alpha\approx \{11,8,3\}$~meV\,nm) or larger ($\alpha\approx \{15,16.5,18.5\}$~meV\,nm)
Rashba coefficients, respectively, as
the initial photon number increases ($N_{ph,\rm init}=\{1,2,3\}$).
The dips are indicated in \fig{current_per_200_p_alpha_nph} by arrows
in the color of the charge current from the left lead, $I^c_L$.
The circularly polarized photon field
has a spin angular momentum, which is proportional to the number of photons in the system.
The spin angular momentum of the photons interacts with the total (orbital 
and spin) angular momentum
of the electrons in the ring.
For a one-dimensional ring of radius $a$ with only Rashba interaction~\cite{2013arXiv1310.5870A} the electron states are fourfold degenerate at
$\alpha=0$ and in general twofold degenerate for $\alpha>0$, as will be explained here in detail.
The eigenstates are
\begin{eqnarray}
 \Psi_{\nu n}^{R}(\varphi)&=&\left(\begin{array}{c} \Psi_{\nu n}^{R}(\varphi,\uparrow)\\ \Psi_{\nu n}^{R}(\varphi,\downarrow)\end{array}\right) \nonumber \\
 &=&\frac{\exp(in\varphi)}{\sqrt{2\pi a}}\left(\begin{array}{c} A_{\nu,1}^{R} \\ A_{\nu,2}^{R}\exp(i\varphi) \end{array}\right) \label{Rwfc}
\end{eqnarray}
with the $2\times 2$ coefficient matrix
\begin{equation}
 A^{R}=\left(\begin{array}{cc} A_{\nu,1}^{R} & A_{\nu,2}^{R} \end{array}\right)=\left(\begin{array}{cc} \cos\left(\frac{\theta_{R}}{2}\right) & \sin\left(\frac{\theta_{R}}{2}\right) \\
 \sin\left(\frac{\theta_{R}}{2}\right) & -\cos\left(\frac{\theta_{R}}{2}\right)
 \end{array}\right), \label{Rcoef}
\end{equation}
\begin{equation}
 \tan\left(\frac{\theta_{R}}{2}\right)=\frac{1-\sqrt{1+x_R^2}}{x_R} \label{Rtantheta}
\end{equation}
and the dimensionless Rashba parameter, $x_R$ (and Dresselhaus parameter, $x_D$) is defined by
\begin{equation}
 \left(\begin{array}{c}x_{R}\\x_{D}\end{array}\right) :=\frac{2m^{*}a}{\hbar^2}\left(\begin{array}{c}\alpha \\ \beta \end{array}\right). \label{defxR}
\end{equation}
We remind the reader that $a$ is the ring radius. We call $n$ the total angular momentum quantum number and 
$\nu=\pm 1$ the spin quantum number (the latter according to the cardinality of the set of possible values). 
One can show that $n$ describes indeed the total (i.e.\ spin and orbital) angular momentum:
\begin{eqnarray}
 \bra{\nu n}\hat{J}_z\ket{\nu n}&=&\bra{\nu n}\hat{L}_z+S_z\ket{\nu n}\nonumber \\
 &=&\bra{\nu n}\frac{\hbar}{i}\frac{\partial}{\partial \varphi}+\frac{\hbar}{2}\sigma_z\ket{\nu n}\nonumber \\
 &=&\hbar\left(n+\frac{1}{2}\right). \label{totangmom}
\end{eqnarray}
In the Dresselhaus case, $\bra{\nu n}\hat{J}_z\ket{\nu n}$ in \eq{totangmom} would depend on $n$ \textit{and}
$x_D$ since $[\hat{J}_z, \hat{H}_D]\neq 0$.
We note in passing that $[\hat{J}_z, \hat{H}_R]=0$
such that, in the Rashba case, $J_z=n+1/2$ is indeed a ``good'' quantum number (constant).
Furthermore, one can define a quantum number
\begin{equation}
 m:=\nu\left(n+\frac{1}{2}\right)+\frac{1}{2}=\nu\frac{\bra{\nu n}\hat{J}_z\ket{\nu n}}{\hbar} +\frac{1}{2}.
\end{equation}
While the exact physical meaning of $m$ and $\nu$ remains unclear, 
these quantum numbers are convenient
to describe the degeneracies in the 1D Rashba ring.
(We assume that they both contain angular
momentum and therefore reflect the spin angular momentum of light.)
States with the same $m$, but different $\nu$
are degenerate for all $\alpha$. Additional degeneracies in $m$ appear at $\alpha=0$ and further
single points, where the states are four fold degenerate in total. 

\begin{figure}[htbq] %bb=1 71 675 355,clip
       \begin{center}
       \includegraphics[width=0.30\textwidth]{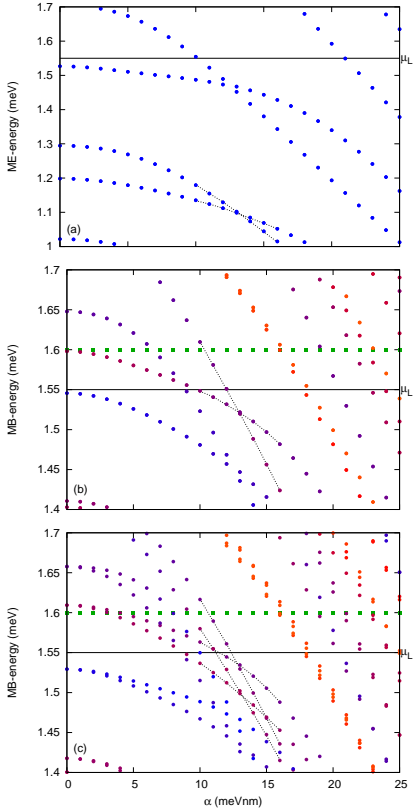}%var width=0.34\textwidth
       \end{center}
       \caption{(Color online) Many-electron (ME) or many-body (MB) energy spectrum of the system Hamiltonian \eq{H^S}
       versus the Rashba coefficient $\alpha$ (a)
       without photon cavity, (b) with $x$-polarized photon field
       and (c) with RH circularly polarized photon field. 
       The states are differentiated according to their electron content $N_e$
       by the shape of the dots
       and according to their fractional photon content $N_{ph}$ by their color.
       Zero-electron states ($N_e=0$, 0ES) are represented as filled squares 
       (the green color means that $N_{ph}=4$) and
       single electron states ($N_e=1$, SES) as filled circles
       with the continuous color spectrum from blue over red to yellow
       corresponding to the range $N_{ph}\in [0,3]$.
       The chemical potential $\mu_L$ of the left lead is shown by a solid black line. 
       The mostly occupied states, which contribute to the current dips, are connected by black dotted lines
       close to their crossings as a guide to the eye. 
       Note that the spectra are shown for different energy ranges.}
       \label{MB_spec_valpha_gluggi}
\end{figure}

How can the degeneracy in $\nu$ be lifted?
As said before, most likely, the spin quantum number $\nu$ contains inherently also orbital angular momentum.
It is though clear, that is contains angular momentum of some kind.
As a consequence, we expect that the circularly polarized photon field
would have a different influence on states of
different $\nu$ (because of the different angular momentum content), 
which would imply that it could lift the degeneracy.
However, we would not expect that the linearly polarized photon field could lift the degeneracy,
as it couples not to the angular momentum.
Only the circularly polarized photon field with a non-vanishing spin angular momentum
can therefore lift the degeneracy in $\nu$.
It is interesting to note that
for the two-dimensional ring without photons
(\fig{MB_spec_valpha_gluggi}(a)) or linearly polarized photons
(\fig{MB_spec_valpha_gluggi}(b)), the states are in general double degenerate
for all values of $\alpha$ except single crossing points. The $m$-degeneracy at $\alpha=0$ 
is split due to the 2D geometry.
The $\nu$-degeneracy and its energy splitting for $\alpha>0$
by the circularly polarized photon field
is here of main interest 
(\fig{MB_spec_valpha_gluggi}). 
We would like to draw the attention of the reader to subtle difference between the 1D and 2D case.
First, we consider the 1D case with two states with different $m$, which are splitted in energy for $\alpha>0$.
Second, we consider the 2D case with circularly polarized photon field with two states with different $\nu$,
which are splitted in energy for $\alpha>0$.
In both cases, we look at the crossings at $\alpha>0$ around the critical values in $\alpha$
corresponding to the AC phase differences $\Delta \Phi=n\pi$ with $n=1,2,\dots$.
Then, the difference can be stated as follows:
in the 1D case one state crosses with a state that is lower, and the other state,
with a state that is higher in energy at $\alpha=0$; in the 2D case the two states cross also with
two other states, but here the latter are degenerate at $\alpha=0$.

To understand the three dips in the non-local (lead-ring) charge current better, we take a look at the MB spectrum. 
Figure \ref{MB_spec_valpha_gluggi} shows the energy spectrum of the central system versus the Rashba
coefficient. We note that \fig{MB_spec_valpha_gluggi}(c) is independent of the handedness
of the circularly polarized photon field.
The state crossing of the mostly occupied states leading to dips in the non-local charge current
are shown by dotted black lines. These states include at least about 50\% of the charge in the central quantum ring system.
The zero-electron states are shown by filled squares and the SES by filled circles.
The photon content, which can be a fractional number due to the light-matter coupling is shown by a continuous range of colors. A photon content of $N_{ph}=0$ is shown in blue, $N_{ph}=1.5$ is shown in red and
$N_{ph}=3$ is shown in yellow. $N_{ph}=4$ is shown in green color.
When we start with one photon initially, $N_{ph,\rm init}=1$, the MB spectrum for the linearly and circularly polarized fields shown
in \fig{MB_spec_valpha_gluggi}(b) and (c) lead to the situation that
the mostly occupied states are states with a photon content $N_{ph}\approx 1$. These states have a color
close to purple or violet red and lie $0.4$~meV higher in the spectrum than the mostly occupied states
without photon cavity (which have of course photon content $N_{ph}=0$), see \fig{MB_spec_valpha_gluggi}(a).

In the case without photons, \fig{MB_spec_valpha_gluggi}(a),
and with linearly polarized photon field, \fig{MB_spec_valpha_gluggi}(b),
the SES are double degenerate, but become split for $\alpha>0$ in the case of circularly polarized photon field, \fig{MB_spec_valpha_gluggi}(c). Consequently,
without photon cavity or for linearly polarized photons, the four crossings of the four mostly occupied states
are at one value of $\alpha=\alpha^c$. With circularly polarized photon field, the four crossings of the four mostly occupied states become to lie at three different
values of $\alpha$.
Two crossings are at the intermediate $\alpha$-value, which is located close to $\alpha^c$,
the third crossing is at $\alpha<\alpha^c$ and the fourth is at $\alpha>\alpha^c$.
The three MB crossing locations in the circularly polarized photon field case 
are the reason for the three dips of the non-local (lead-ring) charge current.
To summarize the essence of \fig{MB_spec_valpha_gluggi}, each Rashba coefficient $\alpha$ with a crossing
of the mostly occupied states
is corresponding to a dip of the non-local charge current.
Without photon cavity or for linearly polarization of the photon field, we have one Rashba coefficient, $\alpha^c$, with crossings and therefore one dip --- for circular polarization, we have
three values of the Rashba coefficient with crossings and therefore three dips.

\begin{figure}[htbq] %bb=1 71 675 355,clip
       \includegraphics[width=0.34\textwidth,angle=-90]{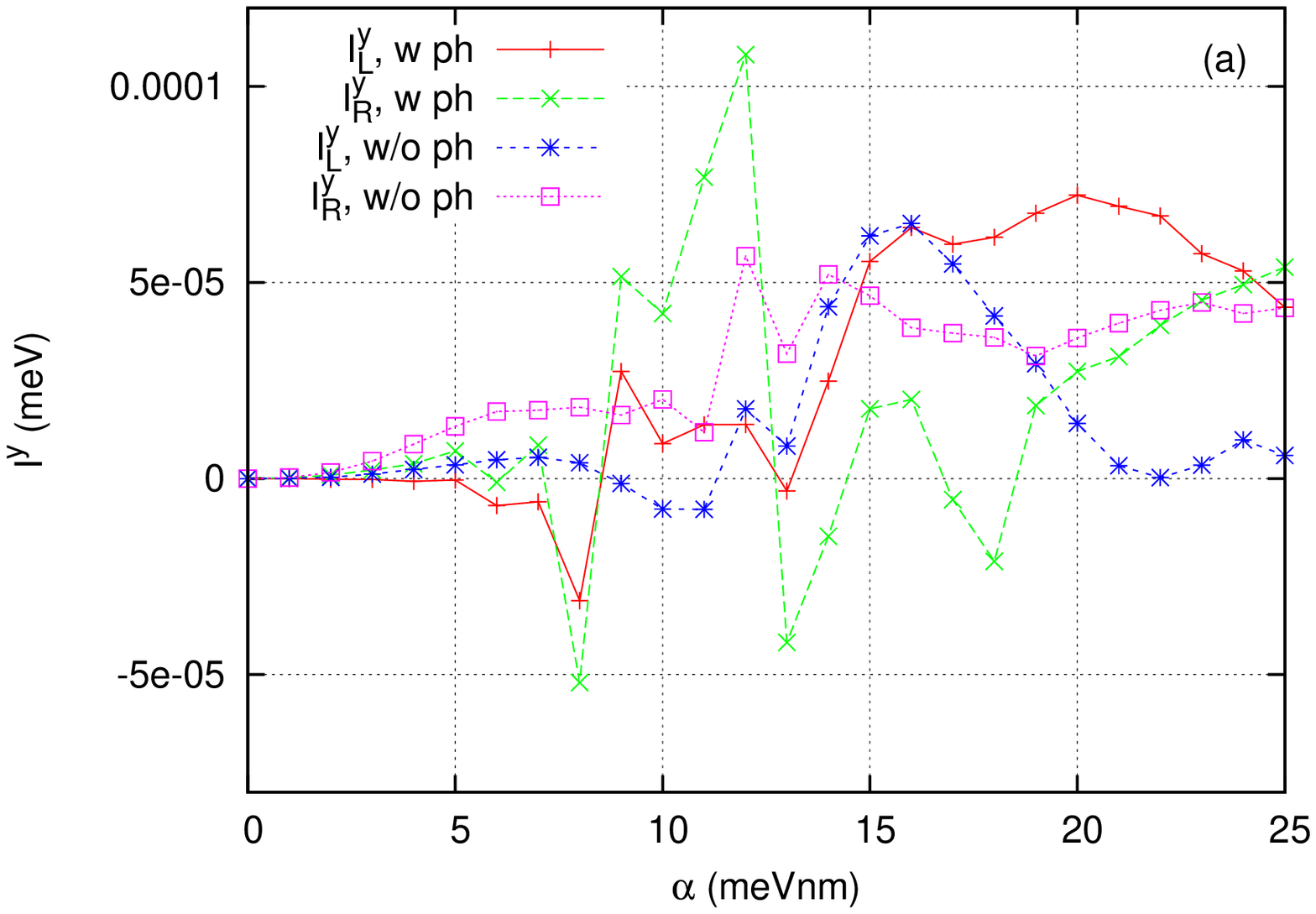}
       \includegraphics[width=0.34\textwidth,angle=-90]{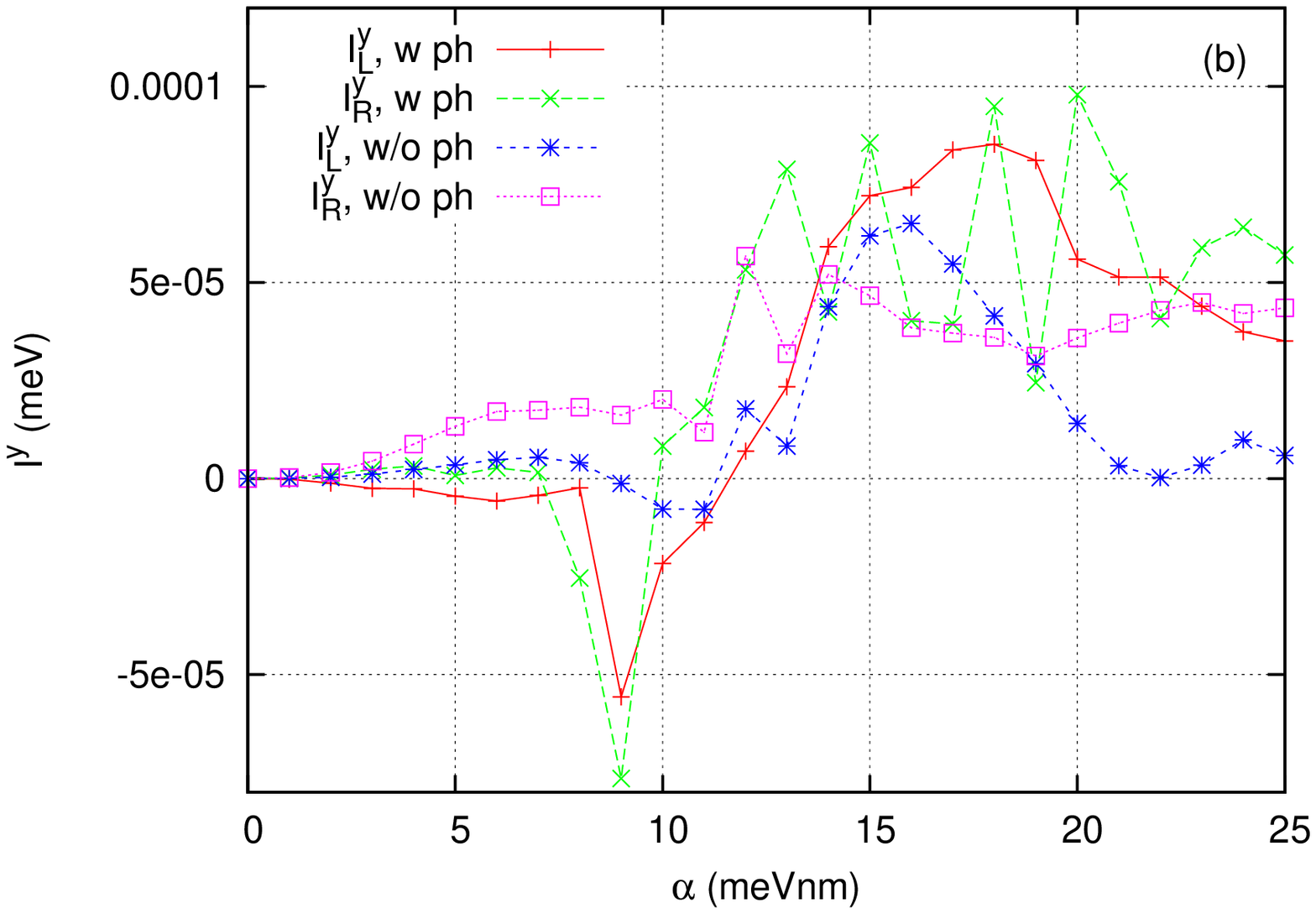}
       \caption{(Color online) Non-local spin polarization currents 
       $I_L^y$ and $I_R^y$ into and out of the central system 
       versus the Rashba coefficient $\alpha$
       averaged over the time interval $[180,220]$~ps
       with (w) (a) $x$-polarized photon field and
       (b) $y$-polarized photon field or
       without (w/o) photon cavity. $I_L^x$, $I_L^z$, $I_R^x$ and $I_R^z$ are not shown
       as they are vanishing.
       The Dresselhaus coefficient $\beta=0$.}
       \label{current_LR_spin_alpha_200}
\end{figure}

Figure \ref{current_LR_spin_alpha_200} shows the non-local (lead-ring) spin polarization currents
for (a) $x$-polarized photon field
and (b) $y$-polarized photon field
from the left lead into the system $\mathbf{I}_L=(I_L^x, I_L^y, I_L^z)$
or from the system to the right lead $\mathbf{I}_R=(I_R^x, I_R^y, I_R^z)$. 
Similarly to the spin polarization, \fig{den_tot_alpha_200}, the $y$-polarized photons 
do \textit{not} induce a non-local current for $S_x$ spin polarization neither from the left nor 
to the right lead. Also the non-local current for $S_z$ spin polarization is vanishing.
In the strong Rashba regime $\alpha \in [19,24]$~meV\,nm, the $S_y$
spin polarization is overall emptied in the system without photon cavity meaning that $I_L^y<I_R^y$, 
however, with $x$-polarized photon field,
we observe in total $S_y$ spin injection ($I_L^y>I_R^y$). 
The $y$-polarized photon field is not in general able
to invert the spin emptying into spin injection in the strong Rashba regime given above.

\subsection{Local photocurrents}

\begin{figure}[htbq] %bb=1 71 675 355,clip    
       \includegraphics[width=0.34\textwidth,angle=-90]{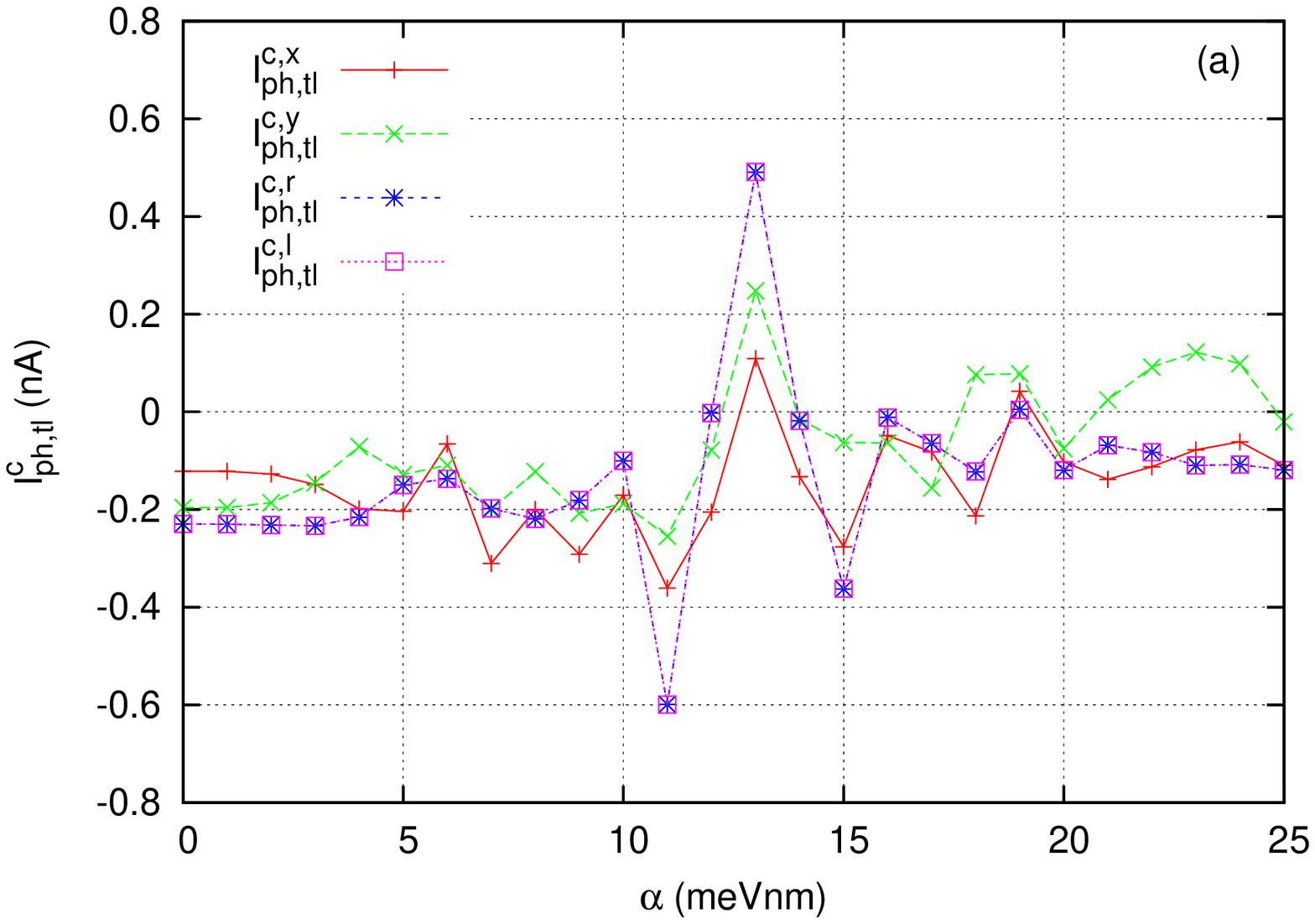}
       \includegraphics[width=0.34\textwidth,angle=-90]{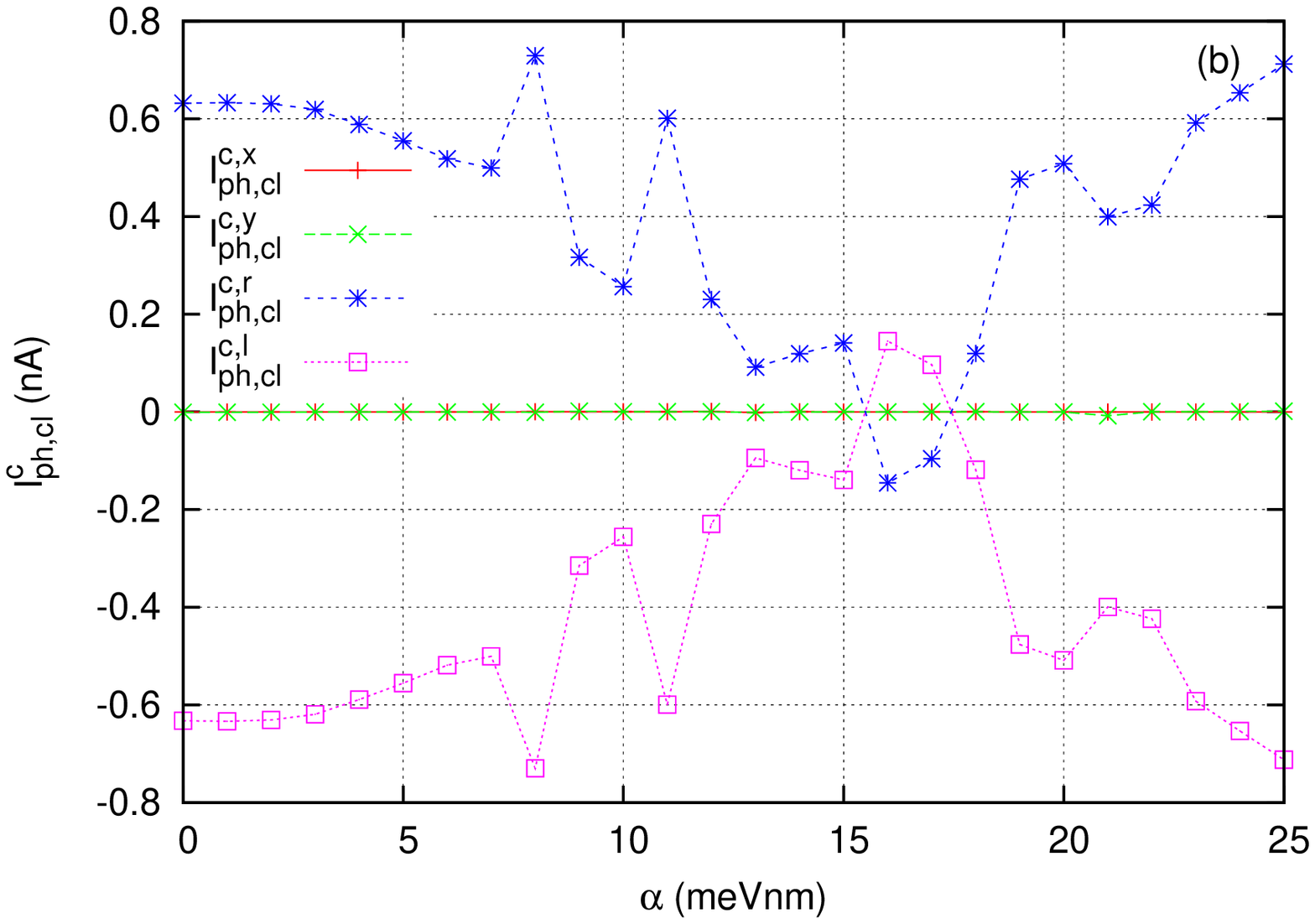}
       \caption{(Color online) (a) TL and
       (b) CL charge photocurrents
       $I^{c,p}_{\rm ph, tl}$ versus the Rashba coefficient $\alpha$
       averaged over the time interval $[180,220]$~ps
       with ($p=x$) $x$-polarized photon field, ($p=y$) $y$-polarized photon field,
       ($p=r$) RH circularly polarized photon field and 
       ($p=l$) LH circularly polarized photon field.
       The Dresselhaus coefficient $\beta=0$.}
       \label{current_arm_tcl_alpha_200}
\end{figure}

Figure \ref{current_arm_tcl_alpha_200} shows the TL and CL charge photocurrents.
The photon cavity reduces in general the TL charge current
(\fig{current_arm_tcl_alpha_200}(a))
(negative photocurrent). 
However, at the destructive AC interference at $\alpha=\alpha^c$, the TL charge current
is enhanced in particular for the circularly polarized photon field.
By the photonic perturbation of the AC phase difference, the electrons can flow
more freely through the ring and the electron
dwell time is reduced.
In the case of the circularly polarized photon field, 
at the smaller and larger values $\alpha\approx \{11,15\}$~meV\,nm,
where the additional non-local (lead-ring) charge current dips appear (\fig{current_per_200_p_alpha_nph}),
the TL charge photocurrent is very negative.
This gives some further evidence
about the photonic nature of the additional non-local charge current dips.
The TL current is independent of the handedness of the circularly polarized photon field
since the ring is otherwise symmetric with respect to the $x$-axis.
The CL charge photocurrent has different sign for RH 
or LH circularly polarized photon field since the circular motion of electrons changes with the
spin angular momentum of the photons in sign (\fig{current_arm_tcl_alpha_200}(b)).
In contrast, the CL charge current
remains uninfluenced by the linearly polarized photon field.
With the aid of the angular motion of electrons induced by the circularly polarized photon field,
the charge flow can be controlled
to pass through the upper or lower ring arm.
Around the destructive AC interference, the CL charge photon current gets suppressed due to
the unfavorable phase relation.
The suppression spans a relatively wide region $\alpha \in [9,23 ]$~meV\,nm when compared
to the non-local charge current dip. 
The CL charge photon current might therefore serve as an alternative tool to detect
AC phase interference phenomena, which minimizes the likelihood to overlook an AC destructive phase
interference because of the narrowness of the non-local charge current dip in the parameter space (consider for example
the dip at $\alpha =\alpha^c$ in \fig{current_per_200_p_alpha_nph}).
We note that the TL and CL charge current is independent from the kind of
SOI, i.e.\ Rashba or Dresselhaus, as it is a spin-independent quantity.

\begin{figure}[htbq] %bb=1 71 675 355,clip
       \begin{center}
       \includegraphics[width=0.34\textwidth,angle=-90]{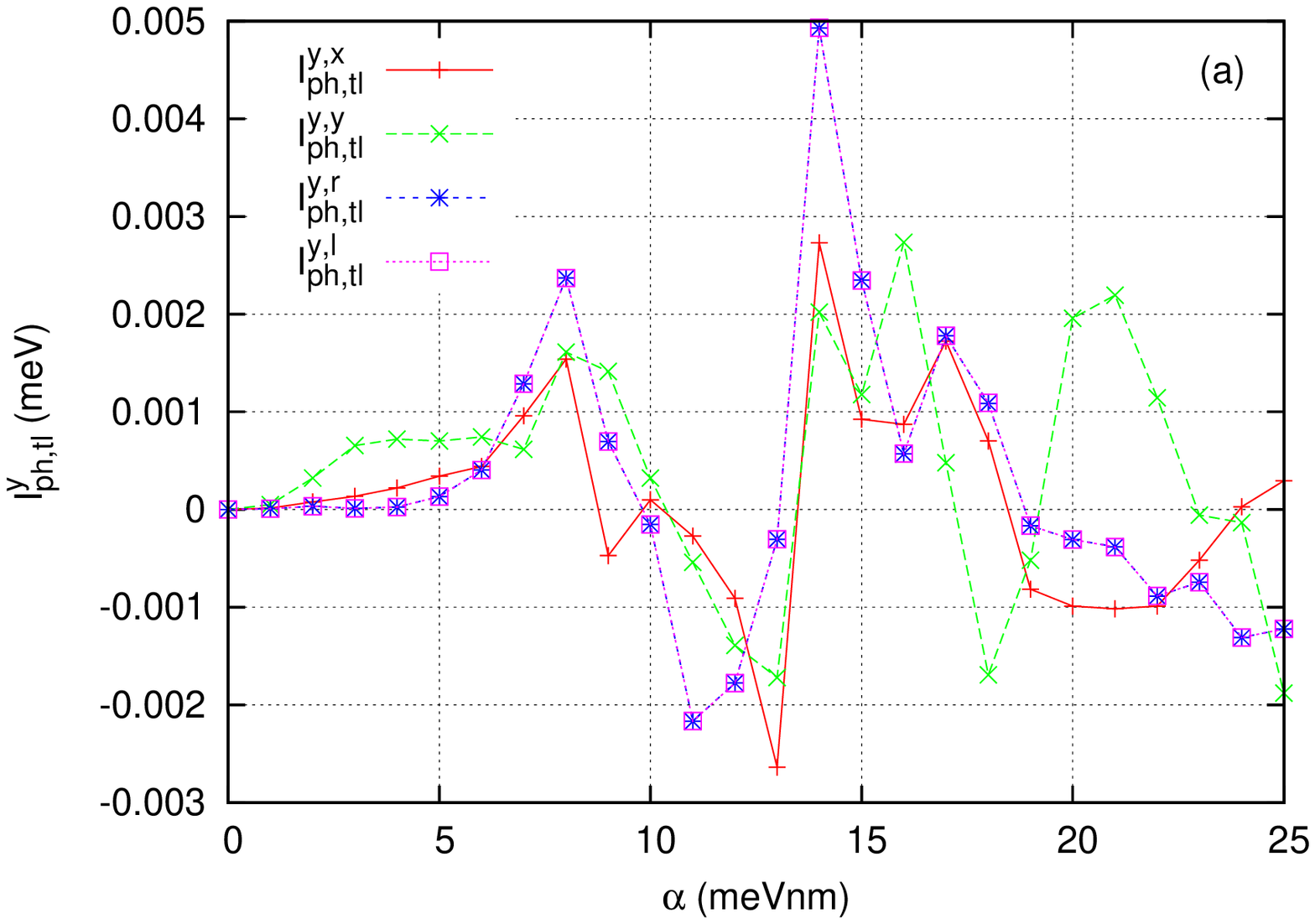}       
       \includegraphics[width=0.34\textwidth,angle=-90]{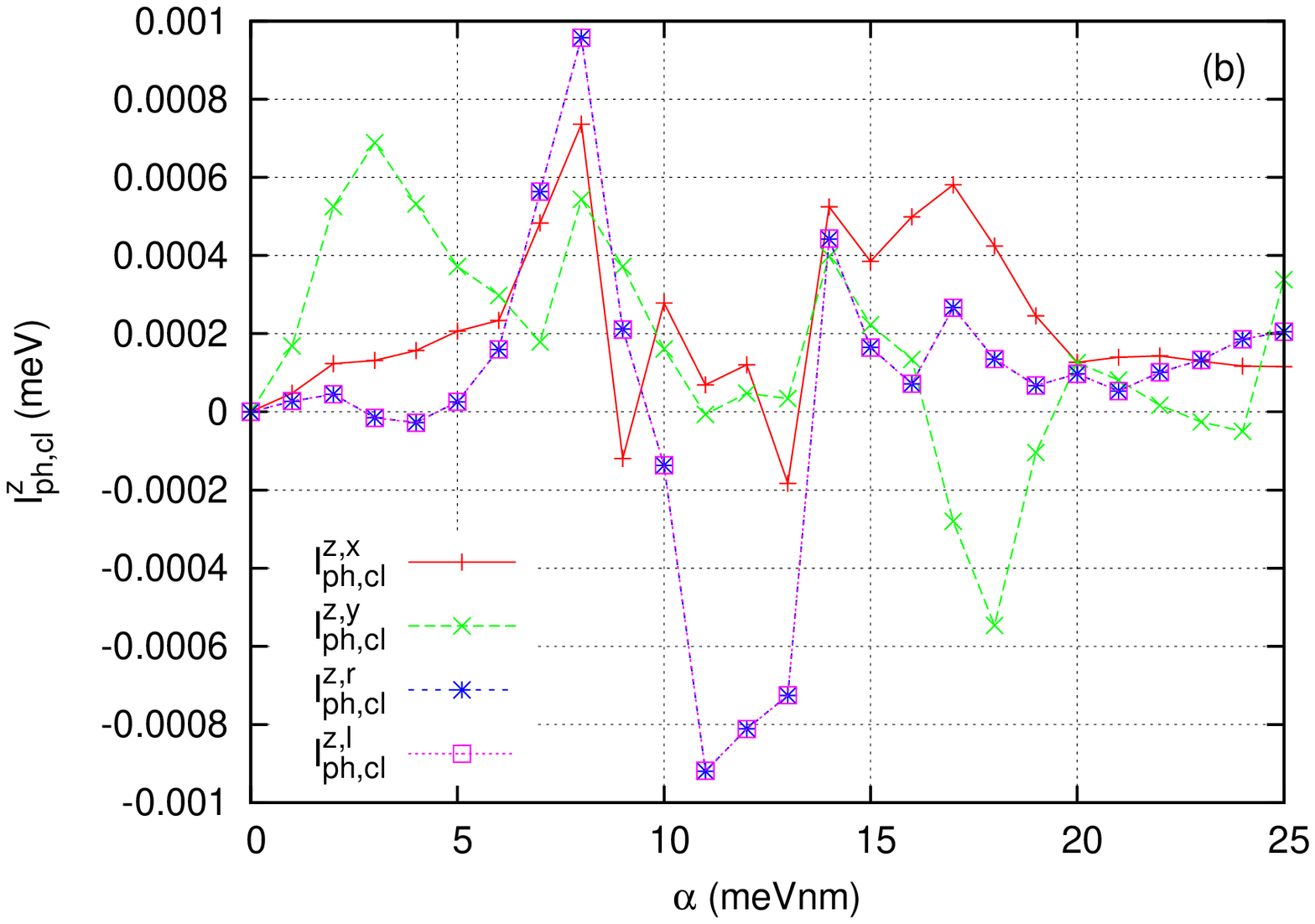}
       \end{center}
       \caption{(Color online) (a) TL spin photocurrents for spin polarization $S_y$,
       $I^{y}_{\rm ph, tl}$ and (b) CL spin photocurrents for spin polarization $S_z$,
       $I^{z}_{\rm ph, cl}$, versus the Rashba coefficient $\alpha$
       averaged over the time interval $[180,220]$~ps
       with ($p=x$) $x$-polarized photon field, ($p=y$) $y$-polarized photon field,
       ($p=r$) RH circularly polarized photon field and 
       ($p=l$) LH circularly polarized photon field.
       The Dresselhaus coefficient $\beta=0$.}
       \label{current_arm_tl_sy_alpha_200}
\end{figure}

Figure \ref{current_arm_tl_sy_alpha_200} shows the TL spin photocurrent for the spin polarization $S_y$
and the CL spin photocurrent for the spin polarization $S_z$.
As opposed to the charge photocurrents, 
the spin photocurrents drop to zero for the Rashba coefficient $\alpha \to 0$ 
(when only a very weak Zeeman term distinguishes the spin).
The influence of the circularly polarized photon field
is strong in a relatively wide range around
the position of the destructive AC phase at $\alpha = \alpha^c$
and weak around the constructive AC phases ($\alpha=0$~meV\,nm and $\alpha\approx 21$~meV\,nm).
For the destructive AC phase, the reduced electron mobility increases the electron dwell time
leading to the strong spin photocurrents.
In general, the influence of the circularly polarized photon field is a bit stronger 
than the influence of the linearly polarized photon field.
We note in passing that the other spin photocurrents, which are not shown, 
$I^{x}_{\rm ph, tl}$, $I^{x}_{\rm ph, cl}$, 
$I^{y}_{\rm ph, cl}$ and $I^{z}_{\rm ph, tl}$, 
are about one order of magnitude smaller than $I^{y}_{\rm ph, tl}$ and $I^{z}_{\rm ph, cl}$.
The local spin polarization currents without photons, $I^{y,0}_{tl}$ and $I^{z,0}_{cl}$, were also much larger
than $I^{x,0}_{tl}$, $I^{x,0}_{cl}$, 
$I^{y,0}_{cl}$ and $I^{z,0}_{tl}$, 
meaning that the photon cavity is not changing the set of major local spin polarization currents.
It is interesting to note that the handedness of the circularly polarized photon field does \textit{not} affect
the major spin photocurrents including even the CL spin photocurrent $I^{z}_{\rm ph, cl}$.

\section{Conclusions} \label{sec5}

The interaction between spin-orbit coupled electrons in a quantum ring interferometer and
a circularly polarized electromagnetic field
shows a variety of interesting effects, which do not appear for linear polarization of the photon field.
The AC phase that controls the transport of electrons
in such a quantum device is influenced by the photons.
We found that the spin polarization in a ring, which is connected to leads 
and mirror symmetric with respect to the transport axis, is perpendicular
to the transport direction. A linearly polarized photon field
with polarization in or perpendicular to the transport direction,
increases only the magnitude of the spin polarization
while keeping the direction of the spin polarization vector uninfluenced.
The spin polarization accumulates to larger magnitudes when 
the transport of electrons is suppressed by a destructive AC phase.
The circularly polarized photon field enhances the spin polarization much more than
the linearly polarized photon field. Furthermore, the spin polarization vector
is no longer bound to a specific direction as the circularly polarized photon field excites 
the orbital angular motion of the electrons around the ring and pronounced vortices
of the charge current density of smaller spatial scale.
The latter show the importance to resolve the finite width of our ring as we did in our model.
The circulation direction of the vortices
is found to depend on the handedness of the photon field 
and the value of the Rashba coefficient $\alpha$ relative to $\alpha^c$.

The charge current from the left lead into the quantum ring device and out to the right lead
shows three AC dips around $\alpha^c$ instead of one for the circularly polarized photon field.
The reason for it is a small splitting of degenerate states by the interaction of the
angular momentum of the electrons and the spin angular momentum of light, which leads to 
MB crossings at three different values of the Rashba coefficient.
The distance in $\alpha$ between the dips increases with the number of photons in the system
due to the larger spin angular momentum of light. 
The charge photocurrent from the left to the right side
of the quantum ring is usually negative meaning that the photon cavity
suppresses the charge transport thus increasing the device resistance
(except close to $\alpha^c$, where the AC phase interference is destructive).
The circulating part of the charge photocurrent can only be excited by the
circularly polarized photon field. The handedness of the circulation depends on the
handedness of the light. This way, it is possible to confine the charge transport
through the ring to one ring arm (upper or lower). The circular charge photocurrent
is suppressed in a wide range of the Rashba coefficient around $\alpha=\alpha^c$ and
might therefore serve as a reliable quantity to detect destructive AC phases.
The spin photocurrents are especially strong around $\alpha=\alpha^c$ (due to the
longer electron dwell time) and for circular polarization (for geometrical reasons). The handedness of the
light does not influence the spin polarization current including the current for $S_z$ spin polarization,
which circulates around the ring.

In summary, strong spin polarization, 
spin photocurrents and charge current vortices as well as
splitting of the AC charge current dip into three dips and control
over the local charge flow through the ring arms are important effects
that only appear for circularly polarized photon field.
These effects are crucial to know about
for the development of spin-optoelectronic quantum devices
in the field of quantum information processing.
For instance, interest might arise to build a spintronic device, which
breaks (blocks) an electrical circuit if the gate voltage is not precisely equal to
a specific, sharply defined critical value,
which corresponds to the magnitude of the electric field leading
to a destructive AC phase interference in a ring interferometer. 
The critical gate potential of the quantum switch could be adjusted
by variation of the ring radius~\cite{2013arXiv1310.5870A}.
A possible experimental approach to determine the ring radius
would be to measure the circular charge current around the ring
(for example indirectly by its induced magnetic field)
that is caused by a circularly polarized cavity photon field.
We predict that this approach is better than
direct resistance measurements of the quantum switch
without the photon cavity. 
This is because the data that a certain number of measurements
with the circularly polarized cavity photon field yields are more relevant
for suggesting the proper ring radius
due to the broadness 
of the corresponding Aharonov-Casher feature in the Rashba coefficient.

\section*{Acknowledgments}
      This work was financially supported by the Icelandic Research
      and Instruments Funds, the Research Fund of the University of Iceland, and the
      National Science Council of Taiwan under contract
      No.\ NSC100-2112-M-239-001-MY3. We acknowledge also support 
      from the computational facilities of the Nordic High Performance Computing (NHPC).

\appendix

\section{Parameters used for the numerical results}\label{parnr}

We assume GaAs-based
material with electron effective mass $m^*=0.067m_e$ and background
relative dielectric constant $\kappa = 12.4$. 
As stated earlier, the Rashba coefficient $\alpha$ can be tuned by
changing the magnitude of an electric field, which is perpendicular
to the plane containing the quantum ring structure. 
The range of $\alpha$ investigated in this paper is about one order of
magnitude larger than typical values of $\alpha$ for GaAs.
However, we point out that the predicted features 
are at fixed positions in $x_R$ (\eq{defxR}) and not in $\alpha$.
Therefore, by increasing the ring radius, experiments
could be performed in a smaller range of the Rashba coefficient 
if it seems difficult to increase the electric field sufficiently
by a gate. For our numerical calculations, it is inconvenient
to increase the ring radius further, as we would have to consider
a larger number of many-body states to get converged results.
With the state of the art computational facilities, however, we are
limited to about $200$ many-body states for our numerically exact approach.
Alternatively, other materials as InAs could be used, for which the
the Rashba coefficient is about one order of magnitude larger~\cite{InAs}.
The Dresselhaus coefficient $\beta$ is determined by the bulk properties
of the material and could only be changed by
using a different material. The value for GaAs would be
$\beta=3.0$~meV\,nm. When using $\beta>0$, we would also
decrease slightly the $\alpha$-range for which our features due to the
destructive AC phase appear. However, our features would become 
more complex when both Rashba and Dresselhaus
spin-orbit interaction are present~\cite{PhysRevB.72.165336}.

We consider a single photon cavity mode
with fixed photon excitation energy $\hbar\omega = 0.4$~meV. The
electron-photon coupling constant in the central system $g^{EM} =
0.1$~meV. The temperature of
the reservoirs is $T = 0.5$~K. The chemical potentials
in the leads are $\mu_L=1.55$~meV and $\mu_R=0.7$~meV leading to a
source-drain electrical bias window $\Delta \mu = 0.85$~meV.

A very small external uniform magnetic field $B=10^{-5}$~T is applied through
the central ring system and the lead reservoirs to lift the spin degeneracy
in the numerical calculations. 
The applied magnetic field $B<<B_0=\Phi_0/A\approx 0.2$~T is order of magnitudes
outside the AB regime. The two-dimensional magnetic length would be very large: $l =
[c\hbar/(eB)]^{1/2} = 8.12$~$\mu$m.  
However, the parabolic confinement of the ring system
in $y$-direction leads to the much shorter magnetic
length scale
\begin{eqnarray}
a_w &= \left(\frac{\hbar}{m^*\Omega_0}\right)^{1/2}
    \frac{1}{\sqrt[4]{1+[eB/(m^*c\Omega_0)]^2}}\nonumber\\
    &\approx \left(\frac{\hbar}{m^*\Omega_0}\right)^{1/2} = 33.7\ {\rm nm}.
\end{eqnarray}

To model the coupling between the system and the leads, we let the affinity
constant $\Delta_E^{l} = 0.25$~meV to be close 
to the characteristic electronic excitation energy in $x$-direction. In
addition, we let the contact region parameters for lead $l\in\{L,R\}$ in $x$-
and $y$-direction be $\delta^l_x = \delta^l_y= 4.39\times
10^{-4}$~$\rm{nm}^{-2}$.  The system-lead coupling strength $g_0^l =
1.371\times 10^{-3}$~${\rm meV}/{\rm nm}^{3/2}$.
Before switching on the system-lead coupling at $t=0$ with
the time-scale $(\alpha^{l})^{-1}=3.291$~\textrm{ps}, we assume the central
system to be in the pure state with electron occupation
number $N_{e,\rm{init}}=0$ and --- unless otherwise stated --- photon occupation
number $N_{ph,\rm{init}}=1$.
The SES charging time-scale $\tau_{\rm SES}\approx 30$~\textrm{ps},
and the two-electron state (2ES) charging time-scale $\tau_{\rm 2ES}
\gg 200$~\textrm{ps}, 
which is described in the sequential tunneling regime.
We study the non-equilibrium transport properties around $t = 200$~ps,
when the system has not yet reached a steady state.
Some dynamical observables are averaged over the time interval $[180,220]$~ps
to give a more representative picture in the transient regime.
The charge in the quantum ring system at $t = 200$~ps
is typically of the order of $Q(t=200\rm{ps})\approx 1e$.

\section{Operators for the charge and spin polarization density and charge and spin polarization current density}\label{chspdcdop}

The charge density operator
\begin{equation}
 \hat{n}^{c}(\mathbf{r})=e\hat{\mathbf{\Psi}}^{\dagger}(\mathbf{r})\hat{\mathbf{\Psi}}(\mathbf{r}) \label{ncq}
\end{equation}
and the spin polarization density operator for spin polarization $S_i$
\begin{equation}
 \hat{n}^{i}(\mathbf{r})=\frac{\hbar}{2}\hat{\mathbf{\Psi}}^{\dagger}(\mathbf{r})\sigma_{i}\hat{\mathbf{\Psi}}(\mathbf{r}). \label{niq}
\end{equation}
The component labeled with $j\in \{x,y\}$ of the charge current density operator
\begin{eqnarray}
\hat{j}^{c}_{j}(\mathbf{r})&=&\frac{e\hbar}{2m^{*}i}\left[\hat{\mathbf{\Psi}}^{\dagger}(\mathbf{r})\nabla_{j} \hat{\mathbf{\Psi}}(\mathbf{r}) - \left(\nabla_{j} \hat{\mathbf{\Psi}}^{\dagger}(\mathbf{r})\right)\hat{\mathbf{\Psi}}(\mathbf{r}) \right]  \nonumber \\
&& + \frac{e^2}{m^{*}c}\hat{A}_{j}(\mathbf{r}) \hat{\mathbf{\Psi}}^{\dagger}(\mathbf{r})\hat{\mathbf{\Psi}}(\mathbf{r})\nonumber\\
&&+\frac{e}{\hbar}\hat{\mathbf{\Psi}}^{\dagger}(\mathbf{r}) (\beta \sigma_x-\alpha \sigma_y) \hat{\mathbf{\Psi}}(\mathbf{r}) \delta_{x,j}\nonumber \\
&&+\frac{e}{\hbar}\hat{\mathbf{\Psi}}^{\dagger}(\mathbf{r}) 
(\alpha \sigma_x- \beta \sigma_y)\hat{\mathbf{\Psi}}(\mathbf{r})\delta_{y,j} \label{jcq}
\end{eqnarray}
with the space-dependent vector potential including the static magnetic field and cavity photon field part
\begin{equation}
 \hat{\mathbf{A}}(\mathbf{r})=\mathbf{A}(\mathbf{r}) +
 \hat{\mathbf{A}}^{\mathrm{ph}}(\mathbf{r}).
\end{equation}
The current density operator for the $j$-component and $S_x$ spin polarization
\begin{eqnarray}
\hat{j}^{x}_{j}(\mathbf{r})&=&\frac{\hbar^2}{4m^{*}i}\left[\hat{\mathbf{\Psi}}^{\dagger}(\mathbf{r})\sigma_{x}\nabla_{j} \hat{\mathbf{\Psi}}(\mathbf{r}) \right. \nonumber \\
&& - \left. \left(\nabla_{j} \hat{\mathbf{\Psi}}^{\dagger}(\mathbf{r})\right)\sigma_{x}\hat{\mathbf{\Psi}}(\mathbf{r}) \right]  \nonumber \\
&& + \frac{e \hbar}{2 m^{*}c}\hat{A}_{j}(\mathbf{r}) \hat{\mathbf{\Psi}}^{\dagger}(\mathbf{r})\sigma_x\hat{\mathbf{\Psi}}(\mathbf{r})\nonumber\\
&&+\frac{\beta \delta_{x,j}+\alpha\delta_{y,j}}{2}\hat{\mathbf{\Psi}}^{\dagger}(\mathbf{r})\hat{\mathbf{\Psi}}(\mathbf{r}). \label{jsxq}
\end{eqnarray}
the current density operator for $S_y$ spin polarization
\begin{eqnarray}
\hat{j}^{y}_{j}(\mathbf{r})&=&\frac{\hbar^2}{4m^{*}i}\left[\hat{\mathbf{\Psi}}^{\dagger}(\mathbf{r})\sigma_{y}\nabla_{j} \hat{\mathbf{\Psi}}(\mathbf{r}) \right. \nonumber \\
&& - \left.\left(\nabla_{j} \hat{\mathbf{\Psi}}^{\dagger}(\mathbf{r})\right)\sigma_{y}\hat{\mathbf{\Psi}}(\mathbf{r}) \right]  \nonumber \\
&& + \frac{e \hbar}{2 m^{*}c}\hat{A}_{j}(\mathbf{r}) \hat{\mathbf{\Psi}}^{\dagger}(\mathbf{r})\sigma_y\hat{\mathbf{\Psi}}(\mathbf{r})\nonumber\\
&&-\frac{\alpha \delta_{x,j}+\beta\delta_{y,j}}{2}\hat{\mathbf{\Psi}}^{\dagger}(\mathbf{r})\hat{\mathbf{\Psi}}(\mathbf{r}). \label{jsyq}
\end{eqnarray}
and $S_z$ spin polarization
\begin{eqnarray}
\hat{j}^{z}_{j}(\mathbf{r})&=&\frac{\hbar^2}{4m^{*}i}\left[\hat{\mathbf{\Psi}}^{\dagger}(\mathbf{r})\sigma_{z}\nabla_{j} \hat{\mathbf{\Psi}}(\mathbf{r}) \right. \nonumber \\
&& - \left. \left(\nabla_{j} \hat{\mathbf{\Psi}}^{\dagger}(\mathbf{r})\right)\sigma_{z}\hat{\mathbf{\Psi}}(\mathbf{r}) \right]  \nonumber \\
&& + \frac{e\hbar}{2m^{*}c}\hat{A}_{j}(\mathbf{r}) \hat{\mathbf{\Psi}}^{\dagger}(\mathbf{r})\sigma_{z}\hat{\mathbf{\Psi}}(\mathbf{r}). \label{jszq}
\end{eqnarray}

\section*{References}

\bibliographystyle{model1a-num-names}

%-----------------------------------------------------------
%
\end{document}